\documentclass{article}
\usepackage{ijcai22}

\usepackage{times}
\usepackage{soul}
\usepackage{url}
\usepackage[hidelinks]{hyperref}
\usepackage[utf8]{inputenc}
\usepackage[small]{caption}
\usepackage{graphicx}
\usepackage{amsmath}
\usepackage{amsthm}
\usepackage{booktabs}
\usepackage{algorithm}
\usepackage{algorithmic}
\usepackage{amssymb}
\usepackage{pifont}
\usepackage{subfigure}
\usepackage{enumitem}
\usepackage{color}

\newcommand{\nosection}[1]{\vspace{2pt}\noindent\textbf{#1}}
\usepackage{multicol}
\usepackage{multirow}
\newcommand{\model}{LASER}
\newcommand{\Density}{Cohesion}
\newcommand{\density}{cohesion}

\newtheorem{theorem}{Theorem}
\newtheorem{definition}{Definition}

\title{Making Recommender Systems Forget: Learning and Unlearning \\for Erasable Recommendation}

\author{
Yuyuan Li$^1$
\and
Xiaolin Zheng$^1$\and
Chaochao Chen$^1$\footnote{Contact Author}\And
Junlin Liu$^1$
\affiliations
$^1$Zhejiang University\\
\emails
\{11821022, xlzheng, zjuccc, junlin\}@zju.edu.cn,
}

\begin{document}

\maketitle

\begin{abstract}
Privacy laws and regulations enforce data-driven systems, e.g., recommender systems, to erase the data that concern individuals.
As machine learning models potentially memorize the training data, data erasure should also unlearn the data lineage in models, which raises increasing interest in the problem of Machine Unlearning (MU).
However, existing MU methods cannot be directly applied into recommendation.
The basic idea of most recommender systems is collaborative filtering, but existing MU methods ignore the \textit{collaborative information} across users and items. 
In this paper, we propose a general erasable recommendation framework, namely LASER, which consists of Group module and SeqTrain module.  
Firstly, Group module partitions users into balanced groups based on their similarity of collaborative embedding learned via hypergraph.
Then SeqTrain module trains the model sequentially on all groups with curriculum learning.
Both theoretical analysis and experiments on two real-world datasets demonstrate that LASER can not only achieve efficient unlearning, but also outperform the state-of-the-art unlearning framework in terms of model utility.
\end{abstract}

\section{Introduction}\label{sec:intro}

Recommender Systems (RSs) are typically built by analyzing the data collected from users, such as users' ratings on items.
Existing regulations, e.g., the General Data Protection Regulation~\cite{2014gdpr}, enforce the ability of erasing the personal data that concern individuals, which raises the concerns of privacy in RSs.
Due to the fact that machine learning models, which have been ubiquitously applied in RSs, potentially memorize the training data~\cite{bourtoule2021machine}.
Besides simply deleting the target data, data erasure is also required to unlearning the data lineage in RS models.
In general, it is beneficial for both users and recommendation platforms to build an erasable RS that supports unlearning in addition to learning.
On the one hand, an erasable RS can preserve the privacy of users.
On the other hand, it can also enhance recommendation performance by active unlearning.
This is because the performance of RSs is sensitive to the training data which can be easily polluted by accidental mistakes or poisoned by intentional attack~\cite{schafer2007collaborative}.

\begin{figure}[t]
    \centering
    \includegraphics[width=0.8\linewidth]{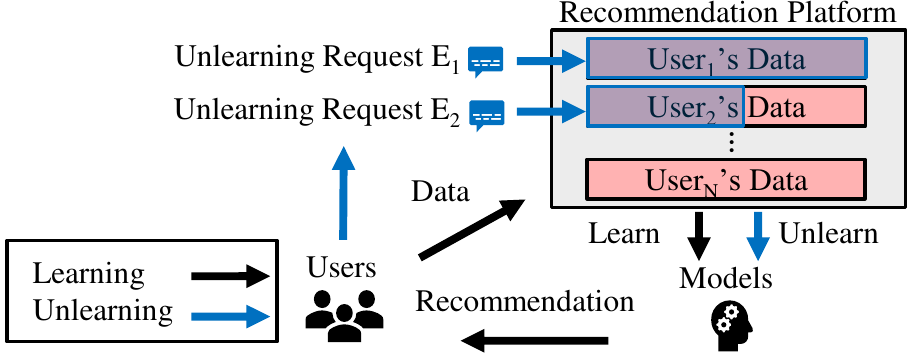}
    \vspace{-0.2cm}
    \caption{A schematic view of learning and unlearning in recommendation.
    }
    \label{fig:schema}
\end{figure}

Figure~\ref{fig:schema} illustrates the schema of learning and unlearning in recommendation.
In the learning process (black arrows), the platform trains recommendation models with the data collected from users.
In the unlearning process (blue arrows), the platform unlearns the target data lineage according to the unlearning requests from users.
There are basically two types of unlearning requests, (i) \textit{user-wise erasure} (E$_1$) which unlearns all of the data from the user(s), and (ii) \textit{data-wise erasure} (E$_2$) which unlearns a portion of user data.
%
%
In order to evaluate unlearning methods, we summarize three goals of unlearning, i.e., G1: \textit{Unlearn Completeness}, G2: \textit{Computational Efficiency}, and G3: \textit{Model Utility}.

Recently, researches have studied the problem of Machine Unlearning (MU) which aims to unlearn the target data lineage in machine learning models.
A straightforward MU method is retraining the model from scratch on the new dataset after deleting the target data, which perfectly achieves G1 and G3, but fails G2.
Due to tremendous computational overhead of recommendation models, efficiency is one of the key concerns for unlearning.
To overcome the issue of inefficiency, two classes of MU methods, i.e., \textit{retrain unlearning} and \textit{reverse unlearning}, have been proposed~\cite{bourtoule2021machine,schelter2021hedgecut,sekhari2021remember}.
Retrain unlearning basically divides the original dataset into subsets and retrains the model on the target subset(s) to reduce computational overhead.
In contrast, reverse unlearning focuses on the target data and executes reverse operations.

Unfortunately, the above MU methods cannot be directly applied into recommendation. 
The reason is that the basic idea of most RSs is Collaborative Filtering (CF)~\cite{shi2014collaborative}, but these MU methods fail to consider the \textit{collaborative information} across users and items. 
On the one hand, retrain unlearning trains each model on a subset of the original dataset, which means that each model can only utilize collaborative information within the subset.
Consequently, retrain unlearning reduces total model utility (G3). 
On the other hand, reverse unlearning only executes reverse operations on the target data ignoring the collaborative effect with associated data. 
Thus, reverse unlearning cannot unlearn with completeness (G1).

In this paper, we focus on user-wise erasure problem and propose \model{}~for it. 
Following the idea of retrain unlearning, \model{} naturally achieves G1.
Specifically, \model{} consists of two modules, i.e, \textit{Group module} and \textit{Sequential Train (SeqTrain) module}, which are constructed not only to achieve efficient unlearning (G2), but also to enhance model utility (G3) by preserving collaborative information.
Firstly, Group module divides the original data into balanced groups for efficient retraining.
These groups are generated based on user similarity, which is measured by the distance of collaborative embedding learned via hypergraph.
Then, SeqTrain module trains the model on all groups sequentially to aggregate collaborative information.
To further improve model performance in such a sequential training manner, SeqTrain module trains the groups in an easy-to-hard order, which imitates the learning style in human curricula.
Our theoretical analysis and empirical study demonstrate that sequential training can improve the performance of models.
Through these two modules, \model{} achieves efficient unlearning and boosts model utility for CF-based recommendation.
We summarize the main contributions of this paper as follows: 
\begin{itemize}[leftmargin=*] \setlength{\itemsep}{-\itemsep}
    \item We focus on user-wise erasure and propose an erasable recommendation framework (\model{}) for CF models.
    \item We encode the high-order collaborations via hypergraph and propose a balanced grouping method to enhance unlearning efficiency.
    \item We propose collaborative \density{} to measure the volume of collaborative information, as well as the learning difficulty. Our theoretical analysis reveals that the easy-to-hard learning order improves model utility.
    \item We conduct extensive experiments on two real-world datasets to demonstrate that \model{} supports efficient unlearning and outperforms the state-of-the-art unlearning framework in model utility.
\end{itemize}

\section{Problem Formulation}

\subsection{Unlearning}

The main problem of building an erasable RS is unlearning and we formulate the process of unlearning as follows:
letting $[N] = \{1, ..., N\}$ denotes a user set, we assume that the recommendation platform learns a model $\mathcal{M}$ on the dataset $\mathcal{D} = \{d_i, i \in [N]\}$ where $d_i$ denotes the data collected from user $i$. 
As it is illustrated by blue arrows in Figure~\ref{fig:schema}, any user $i \in [N]$ can submit an unlearning request $E \subset d_i$ to unlearn the target data.
For user-wise erasure, $E = d_i$, and typically $|\mathcal{D}| \gg |E|$.
In practice, the unlearning requests are often submitted sequentially.
For conciseness, we assume that the platform processes a batch of unlearning requests together. 
Finally, the platform unlearns the model $\mathcal{M}$ based on request $E$, and produces an unlearned model $\mathcal{M}_u$.
Let $\mathcal{M}_u^*$ denote the model that is retrained on $\mathcal{D} \backslash E$ from scratch. 
Although retraining from scratch is extremely inefficient, $\mathcal{M}_u^*$ is the ground-truth unlearned model.
Thus, the three goals of unlearning is equivalent to efficiently producing a model whose distribution is close to that of $\mathcal{M}_u^*$.

\subsection{Goals of Unlearning}\label{sec:goal}


\nosection{G1: Unlearn Completeness.} 
Completely unlearning the target data lineage is one of the most fundamental goals of unlearning, which means fully erasing the influence on model parameters and making it impossible to recover.

\nosection{G2: Unlearn Efficiency.}
Due to the considerable computational overhead of practical recommendation models, unlearn efficiency, especially time efficiency, is an essential goal of unlearning. 

\nosection{G3: Model Utility.}
Although unlearning too much data lineage will inevitably reduce the model utility, an adequate unlearning method should achieve comparable performance with the ground-truth model $\mathcal{M}_u^*$.

\subsection{Choice of Recommendation Models}\label{sec:choice}

Our proposed \model{} can be generalized to most existing CF models. 
%
%
For conciseness, in this paper, we only consider rating data and choose two well-known CF models, i.e., Deep Matrix Factorization (DMF)~\cite{xue2017deep} and Neural Matrix Factorization (NMF)~\cite{he2017neural}. 
Basically, the two models decompose the user-item interaction matrix $R$ into two low-rank embedding matrices, i.e., $\alpha$ and $\beta$, which represent user features and item features respectively. 
The two models predict unknown ratings as follows:
\begin{align}
    & \hat{R}^\text{DMF}_{i, j} = \cos\big(\sigma(\alpha_i), \sigma(\beta_j)\big), \\
    & \hat{R}^\text{NMF}_{i, j} = \sigma\big(F_\text{GMF}(i, j)\oplus F_\text{MLP}(i, j)\big),
\end{align}
where $F_\text{GMF}(i, j) = \sigma(\alpha_i \odot \beta_j)$, $F_\text{MLP}(i, j) = \sigma(\alpha_i \oplus \beta_j)$, $\sigma$ denotes layer operations, $\oplus$ and $\odot$ denote vector concatenation and element-wise product respectively.
Following the original papers, we adopt normalized binary cross entropy and Adam to train the above models.
\section{\model{} Framework}




Generally speaking, LASER follows the idea of retraining unlearning to achieve G1 and G2, and maintains model utility (G3) by preserving collaborative information.

\begin{figure} 
    \centering
    \includegraphics[width=0.85\linewidth]{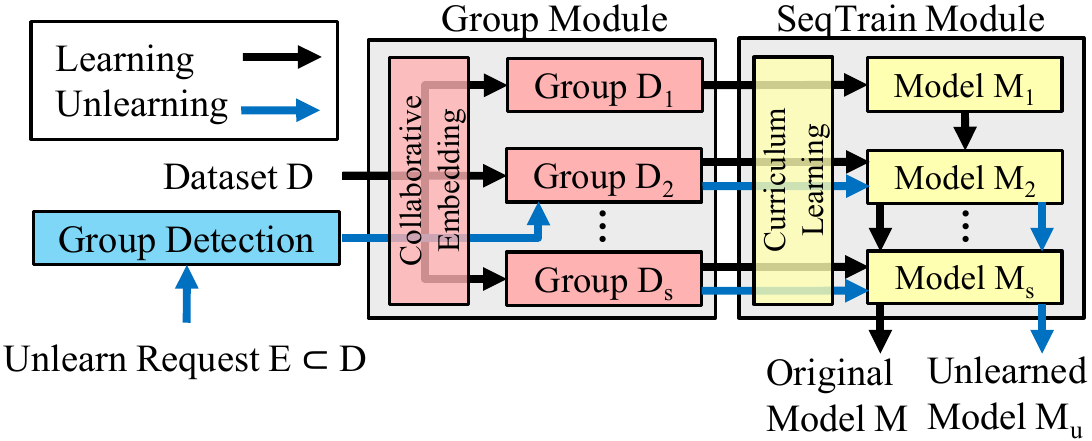}
    \vspace{-0.2cm}
    \caption{Overview of \model{} framework.
    }
    \label{fig:framework}
\end{figure}

\subsection{Overview}

Figure~\ref{fig:framework} illustrates the structure of \model{} framework, which consists of Group module and SeqTrain module.
We introduce the learning and unlearning pipelines as follows:

\nosection{Learning:} 
As the black arrows show in Figure~\ref{fig:framework}, firstly, Group module divides $\mathcal{D}$ into disjoint groups based on the similarity of collaborative embedding.
Secondly, SeqTrain module sorts the groups based on the learning difficulty, then trains the model sequentially in an easy-to-hard order and finally saves these models.
In this way, we can train the model on the whole dataset to preserve collaborative information.

\nosection{Unlearning:}
As the blue arrows show, once an unlearning request $E$ is received, Group module detects in which group $E$ is located and SeqTrain module sequentially retrains the model using the previously saved models.
In this way, we can retrain the model efficiently during unlearning.

\subsection{Group Module}

Group module aims to partition a dataset $\mathcal{D}$ into $S$ groups such that $\bigcap_{i\in[S]} \mathcal{D}_i = \emptyset$ and $\bigcup_{i\in [S]} \mathcal{D}_i = \mathcal{D}$.
This paper focuses on user-wise erasure which means the above dataset partition is equivalent to dividing the user set $[N]$ into $S$ disjoint groups.
Partitioning unlabeled data into groups is a classic unsupervised learning problem named clustering. 
Thus, a straightforward idea is to cluster users based on the observed data. 
There are two grouping principles, i.e., \textit{collaboration} and \textit{balance}, for effective and efficient unlearning in RSs.

%

\subsubsection{Collaboration Principle}

\paragraph{Motivation} Collaboration is of great importance to enhancing RS model utility.
There are two main challenges in collaborative grouping.
Firstly, in the context of RS, directly clustering users suffers from data sparsity.
Since each user only interacts with few items, most elements of the user rating vector, i.e., each row of the user-item interaction matrix $R$, are empty. 
Secondly, the original $R$ or its corresponding bipartite graph cannot sufficiently represent the user-item collaboration, especially high-order relations~\cite{ji2020dual}. 

\begin{figure}[t]
    \centering
    \subfigure[User-item bipartite graph]{
        \label{fig:hg_bi}
        \includegraphics[width=0.43\linewidth]{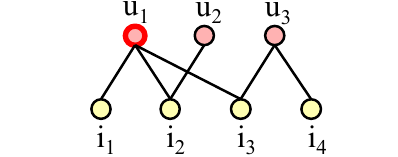}
    }
    \subfigure[High-order relations]{
        \label{fig:hg_h}
        \includegraphics[width=0.4\linewidth]{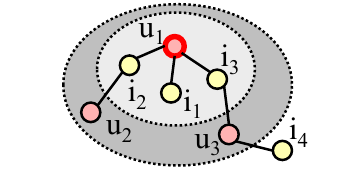}
    }
    \vspace{-0.4cm}
    \caption{An illustration of an user-item bipartite graph and the high-order relations of $u_1$ where $u$ and $i$ denote user and item respectively.}
    \label{fig:hg}
\end{figure}

\paragraph{Method} In order to encode the sparse ratings and enrich the collaborative information, we learn the hidden collaboration via \textit{hypergraph}.
Figure~\ref{fig:hg_bi} depicts a traditional bipartite graph where an edge connects two vertices (a user and an item).
However, Figure~\ref{fig:hg_h} shows the high-order relations of $u_1$ which consists of multiple vertices. 
Hypergraph is a generalized structure for relation modeling, in which a hyperedge connects two or more vertices.
Due to this property, hypergraph can sufficiently model the high-order relations that cannot be directly represented by a traditional graph.

Specifically, we take the following three steps to learn collaborative embedding: 
(i) using $R$ to build the corresponding hypergraph; 
(ii) for each user, performing random walk to obtain its relations, which transforms the task to sequence embedding; 
(iii) applying the sequence embedding technique, e.g., Word2Vec~\cite{church2017word2vec}, to learn the collaborative embedding.
Due to page limit, we summarize the above process in Algorithm~\ref{alg:hg} and present the details in Appendix A.

\begin{algorithm}[tb]
\caption{Collaborative Embedding via Hypergraph}
\label{alg:hg}
\textbf{Input}: User-item interaction matrix: $R$\\
\textbf{Parameter}: Embedding dimension: $M$\\
\textbf{Output}: User embedding matrix: $B \in \mathcal{R}^{N\times M}$\\
\textbf{Procedure}:
\begin{algorithmic}[1] 
\STATE Build hypergraph: $H = BuildHypergraph(R)$;
\STATE Sample user sequence: $F = RandWalk(H)$;
\STATE Generate user embedding: $B = SeqEmbedding(F)$;
\end{algorithmic}
\end{algorithm}

\subsubsection{Balance Principle}

\paragraph{Motivation} After learning the collaborative embedding, the next step is clustering users based on their embeddings.
However, traditional clustering results in unbalanced grouping, which means that the user number in each group, i.e., group size, varies drastically.
We conduct $k$-means~\cite{kanungo2002efficient} clustering on two real-world datasets and report the result in Appendix B.1.
From it, we observe that the grouping distribution is highly unbalanced.
Since we have no prior knowledge about the distribution of unlearning requests, it is better to assume that users submit unlearning requests with equal probability.
The balance principle is making the group size evenly distributed so that it can maximize unlearning efficiency (see Appendix B.2 for theoretical explanation).

\paragraph{Method} To achieve balanced grouping, we propose a general method that can be applied to most existing clustering algorithms, including but not limited to $k$-means clustering, label propagation algorithm~\cite{wang2007label}, and Gaussian mixture models~\cite{reynolds2009gaussian}. 
We present the proposed balanced grouping method in Algorithm~\ref{alg:b}.
To be specific, we set the ceiling size $c_i = \lceil N / S \rceil$ for each group and build a \textit{priority list} $P$ to store the priority of each user-group pair (line 3 in Algorithm~\ref{alg:b}).
For example, in $k$-means clustering, the pair with a smaller Euclidean distance is consider to have a larger priority. 
During every grouping iteration, we allocate the user-group pairs (i) according to their priority, and (ii) if the group has not reached the ceiling size (line 5-8 in Algorithm~\ref{alg:b}).
%
%
We demonstrate balanced $k$-means (BKM) clustering as an example in Appendix B.3.

\begin{algorithm}[tb]
\caption{Balanced Grouping Method}
\label{alg:b}
\textbf{Input}: User embedding matrix: $B$, grouping number: $S$, maximal iteration: $\tau$\\
\textbf{Output}: Group label of $N$ users: $\Lambda = (\lambda_1, .., \lambda_N), \lambda_i \in [S]$
\textbf{Procedure}:
\begin{algorithmic}[1] 
\STATE $t=0$, $c_i = \lceil N / S \rceil$ for $i \in [S]$, randomly allocate $\Lambda$;
\WHILE{True}
    \STATE $P = ComputeSimilarity(B, \Lambda)$;
    \STATE $P = Sort(P, order=descend)$;
    \FOR{$priority(i, j)$ in $P$}
        \IF{$c_j > 0$}
            \STATE $\lambda_i = j$, $c_j = c_j - 1$;
        \ENDIF
    \ENDFOR
    \STATE $t = t + 1$;
\IF{$t > \tau$ or $\Lambda$ do not change}
    \STATE break;
\ENDIF
\ENDWHILE
\end{algorithmic}
\end{algorithm}

\subsection{SeqTrain Module}

Instead of training models on each group in isolation and aggregating them, SeqTrain module trains the model on the whole dataset to preserve collaborative information.
We achieve this by training the model sequentially on all groups, and we further study the effect of training order.

\subsubsection{Training Order}

Based on the theory of Curriculum Learning~\cite{wang2021survey}, training order can make a huge difference in model performance.
We regard each group as a learning task.
Then we aim to train the model from easier tasks to harder ones, which imitates the learning order in human curricula.
A task can be considered as easy if the model has a relatively low loss on it.
The loss can only be obtained either after training or during training.
However, we have to decide the training order before training and cannot regroup the users during training, which means we have to rely on predefined information to measure the learning difficulty of each group.

For CF-based recommendation, collaborative information contributes to better model performance, i.e., lower loss.
It is reasonable to use collaborative information as a difficulty measurer.
In order to measure the volume of collaborative information, we propose the concept of \textit{collaborative \density{}}, which calculates the \density{} of a group. 
The higher the collaborative \density{}, the easier it is for the recommendation model to learn.
The \density{} is defined by the average similarity, i.e., collaboration, of users within the group.
Formally, the collaborative \density{} $\rho$ is computed as:
\begin{equation}
    \rho = \frac{\sum_{x, y \in g} sim(x, y)}{|g|},
\end{equation}
where $g$ denotes a group and $sim(x, y)$ computes the similarity between two users.
In this paper, we set $sim(x, y) = 1/dist(x, y)$, where $dist(x, y)$ denotes Euclidean distance of two users.
%
%
We will empirically study the validity of collaborative \density{} as a difficulty measurer in Section~\ref{sec:ab}.

\subsubsection{Theoretical Analysis} 

Let $\theta$ denote the model parameter vector and $L_\theta(g_i)$ denote the loss of the model when given the data of $i$-th group. 
Adopting the widely used empirical risk minimization framework, we have the empirical loss
\begin{equation}
    \mathcal{L}(\theta) = \hat{\mathbb{E}}[L_\theta] = \frac{1}{S}\sum_{i = 1}^S L_\theta(g_i).
\end{equation}
Minimizing the empirical loss can be regarded as maximizing model utility~\cite{hacohen2019power}, which is defined as:
\begin{align}
    & \mathcal{U}(\theta) = \hat{\mathbb{E}}(U_\theta) = \frac{1}{S}\sum_{i = 1}^S U_\theta(g_i) \triangleq \frac{1}{S}\sum_{i = 1}^S e^{-L_\theta(g_i)}.
\end{align}
As collaborative \density{} $\rho$ indicates the learning difficulty of each group, introducing $\rho$ can be interpreted as providing a Bayesian prior $p$ for model utility. 
\begin{theorem}\label{the:p}
    Given a Bayesian prior $p$ for the model parameters $\theta$, we have:
    \begin{equation}
        \mathcal{U}_p(\theta) = \mathcal{U}(\theta) + \hat{\text{Cov}}[U_\theta, p].
    \end{equation}
\end{theorem}
\begin{proof}
    The proof can be found in Appendix C.
\end{proof}

Theorem~\ref{the:p} reveals that as long as $\hat{\text{Cov}}[U_\theta, p] > 0$, SeqTrain module can improve the original model utility from $\mathcal{U}(\theta)$ to $\mathcal{U}_p(\theta)$.
Our empirical study in Section~\ref{sec:ab} shows that $\rho$ is positively related to $U_\theta$ which confirms that collaborative \density{} is a proper difficult measurer.
\section{Experiments}\label{sec:exp}


\subsection{Dataset}

We evaluate LASER on two publicly accessible datasets: MovieLens 1M (ML)\footnote{https://grouplens.org/datasets/movielens/} and Amazon Digital Music (AM)\footnote{http://jmcauley.ucsd.edu/data/amazon/}. 
The ML and AM datasets are widely used to evaluate CF algorithms~\cite{harper2015movielens,he2016ups}. 
%
%
%
%
We filter out the users and items that have less than 5 interactions. 
We use 90\% of ratings for training and the rest for testing.
Table~\ref{tab:dataset} summarizes the statistics of two datasets.


\begin{table}
\centering
\begin{tabular}{lrrrr}  
\toprule
Dataset & User \#   & Item \#   & Rating \# & Sparsity \\
\midrule
ML      & 6,040     & 3,706     & 1,000,209 & 95.532\%\\
AM      & 478,235   & 266,414    & 836,006   & 99.999\%\\
\bottomrule
\end{tabular}
\caption{Summary of datasets.}
\label{tab:dataset}
\end{table}

\begin{figure}[t]
    \centering
    \subfigure[ML]{
        \label{fig:time_ml1}
        \includegraphics[width=0.2\textwidth]{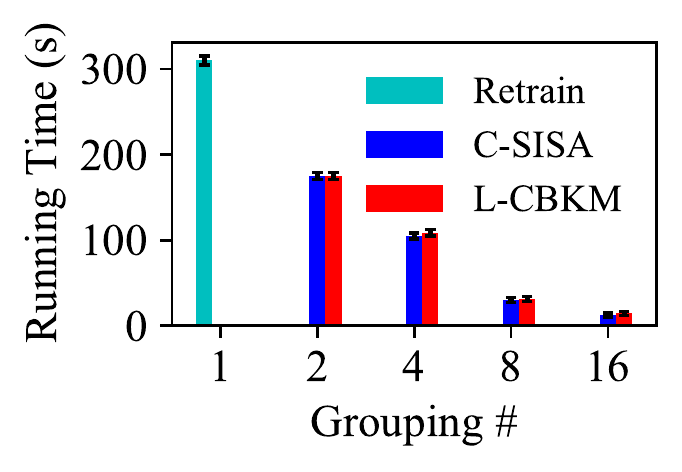}
    }
    \subfigure[AM]{
        \label{fig:time_ah}
        \includegraphics[width=0.2\textwidth]{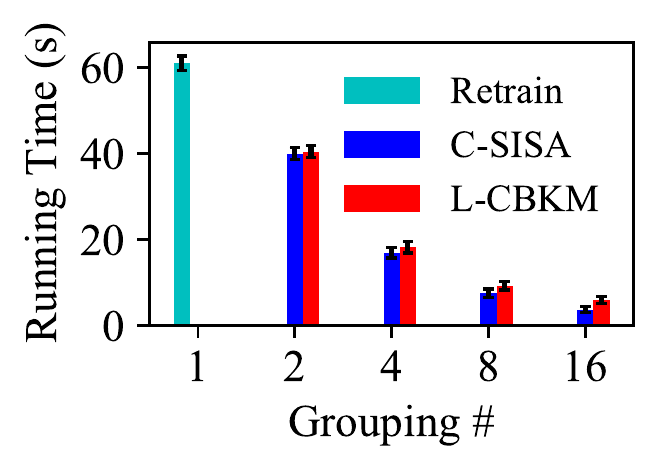}
    }
    \vspace{-0.4cm}
    \caption{Comparison results of running time.}
    \label{fig:time}
\end{figure}

\begin{figure*}[t]
    \centering
    \includegraphics[width=0.5\textwidth]{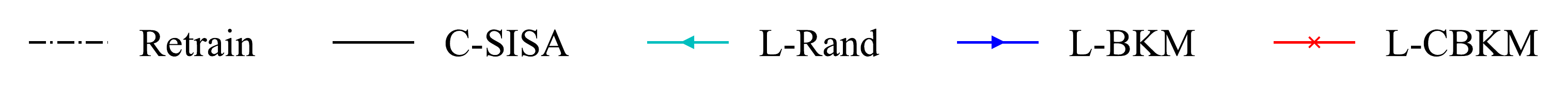}\\
    \vspace{-0.1cm}
    \subfigure[DMF - learn]{
        \includegraphics[width=0.15\textwidth]{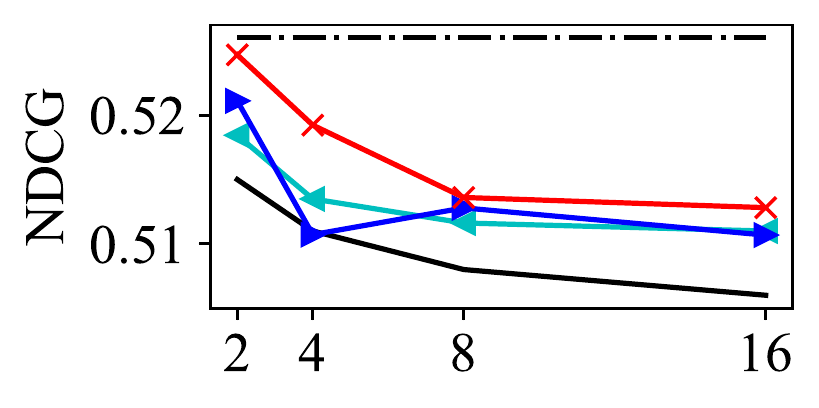}
    }
    \subfigure[NMF - learn]{
        \includegraphics[width=0.15\textwidth]{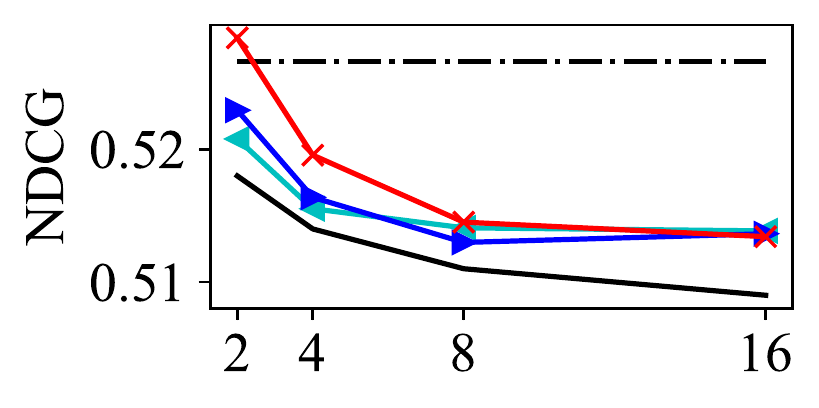}
    }
    \subfigure[DMF - rand@5]{
        \includegraphics[width=0.15\textwidth]{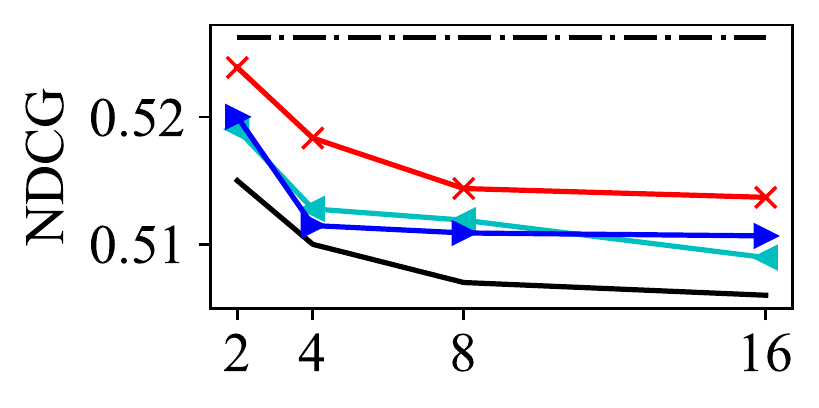}
    }
    \subfigure[NMF - rand@5]{
        \includegraphics[width=0.15\textwidth]{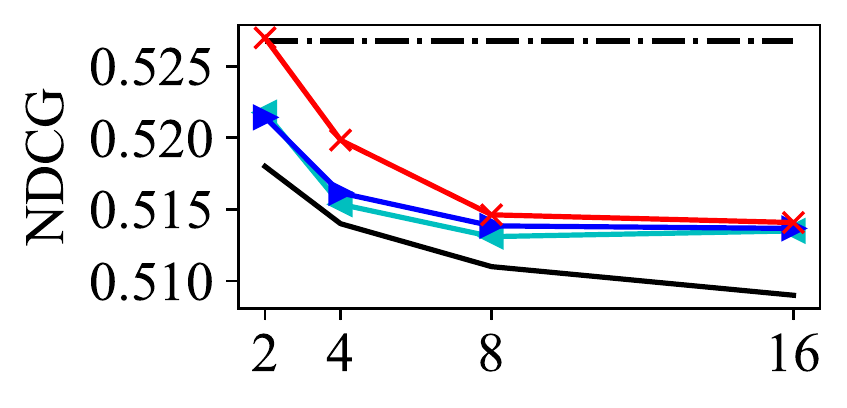}
    }
    \subfigure[DMF - top@5]{
        \includegraphics[width=0.15\textwidth]{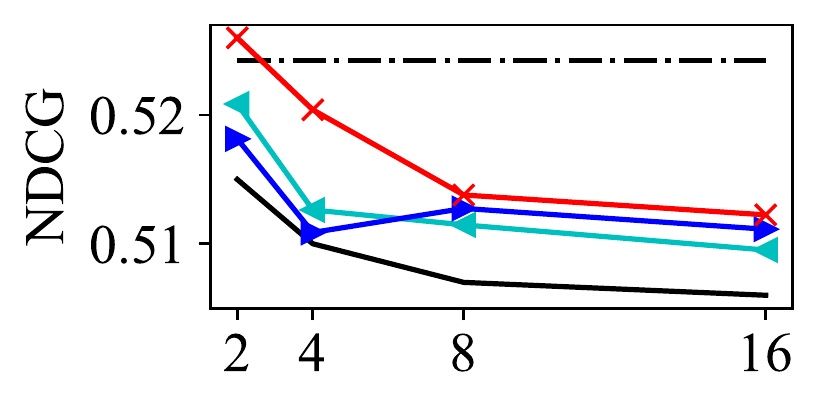}
    }
    \subfigure[NMF - top@5]{
        \includegraphics[width=0.15\textwidth]{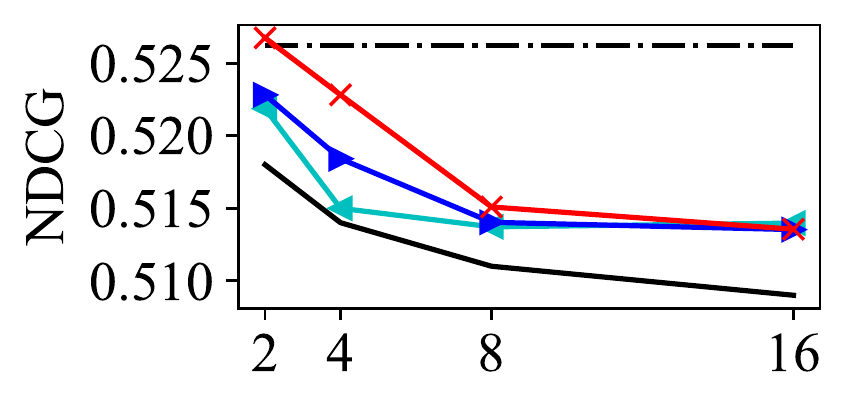}
    }
    \subfigure[DMF - learn]{
        \includegraphics[width=0.15\textwidth]{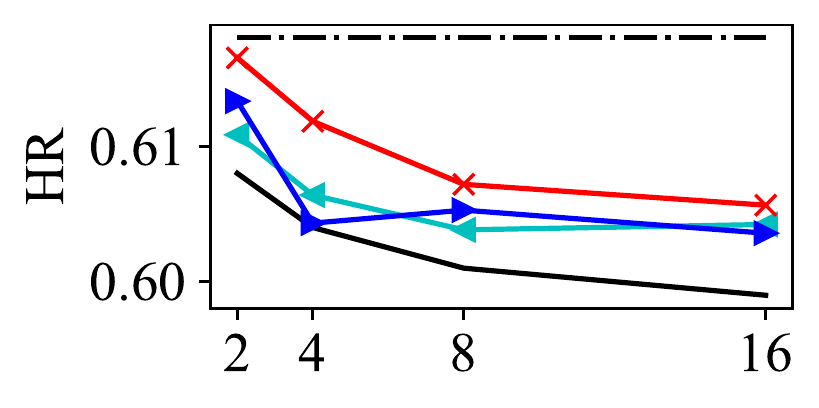}
    }
    \subfigure[NMF - learn]{
        \includegraphics[width=0.15\textwidth]{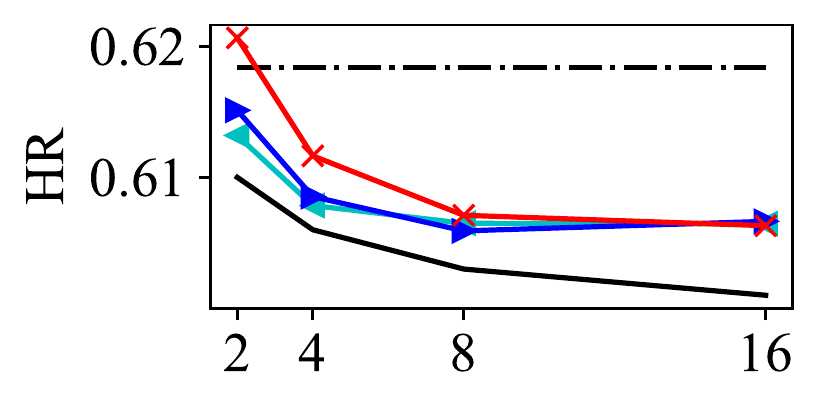}
    }
    \subfigure[DMF - rand@5]{
        \includegraphics[width=0.15\textwidth]{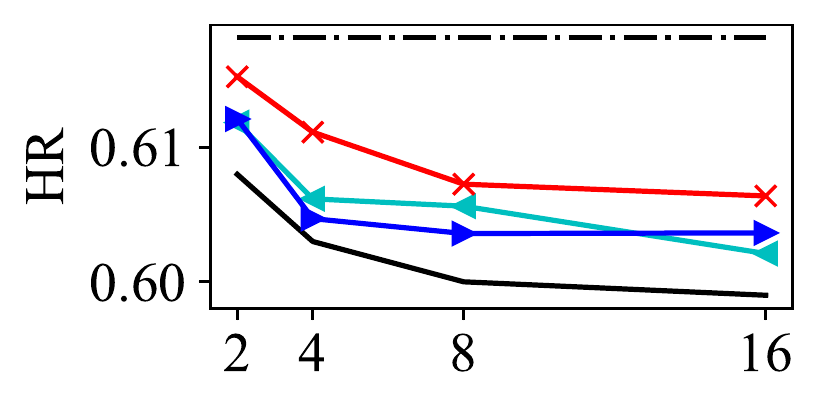}
    }
    \subfigure[NMF - rand@5]{
        \includegraphics[width=0.15\textwidth]{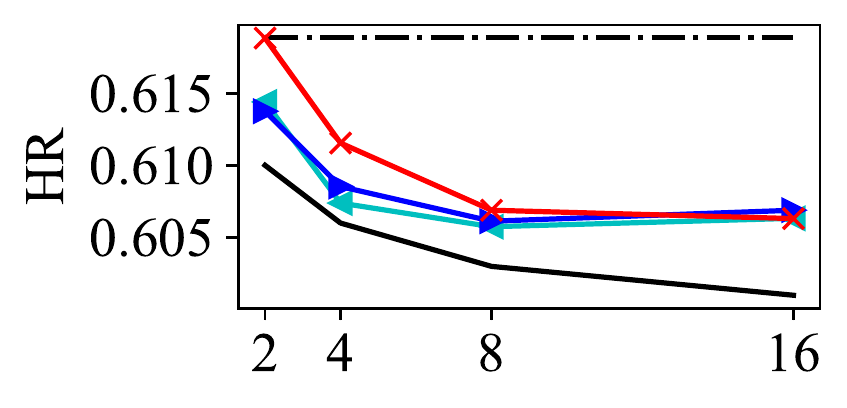}
    }
    \subfigure[DMF - top@5]{
        \includegraphics[width=0.15\textwidth]{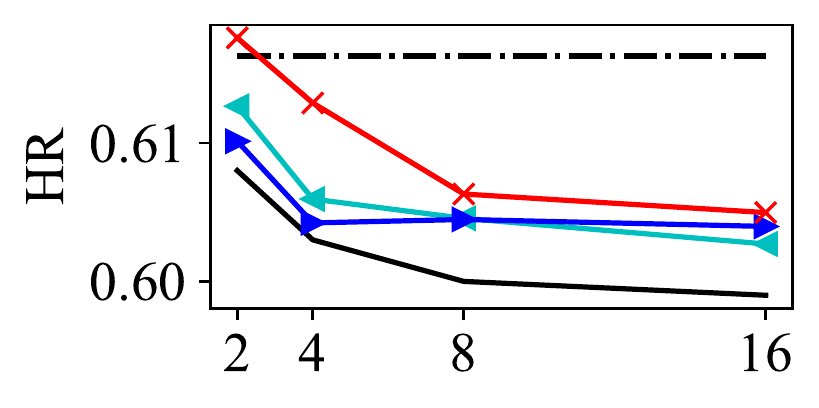}
    }
    \subfigure[NMF - top@5]{
        \includegraphics[width=0.15\textwidth]{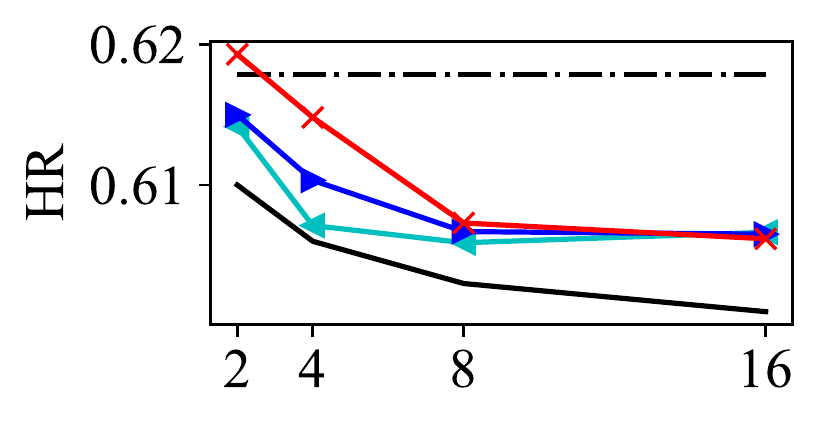}
    }
    \vspace{-0.4cm}
    \caption{NDCG@10 and HR@10 results for utility (G3) study and ablation study (Group module) on ML dataset with std under 1e-3.}
    \label{fig:metric_ml}
\end{figure*}

\begin{figure*}[t]
    \centering
    \includegraphics[width=0.5\textwidth]{fig/metirc_legend.pdf}\\
    \vspace{-0.15cm}
    \subfigure[DMF - learn]{
        \includegraphics[width=0.15\textwidth]{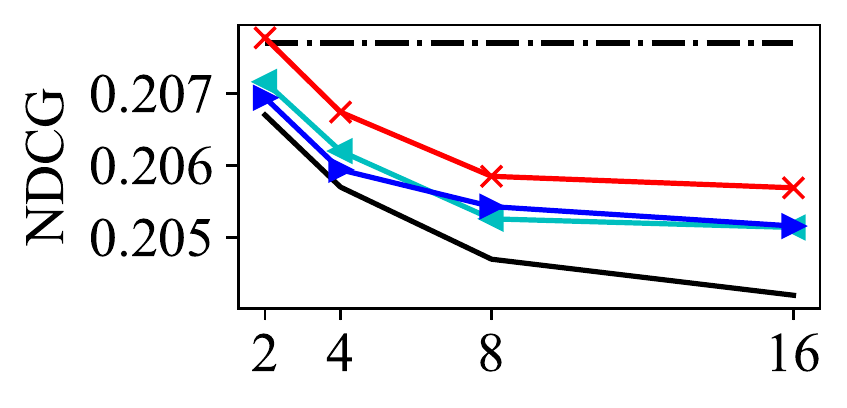}
    }
    \subfigure[NMF - learn]{
        \includegraphics[width=0.15\textwidth]{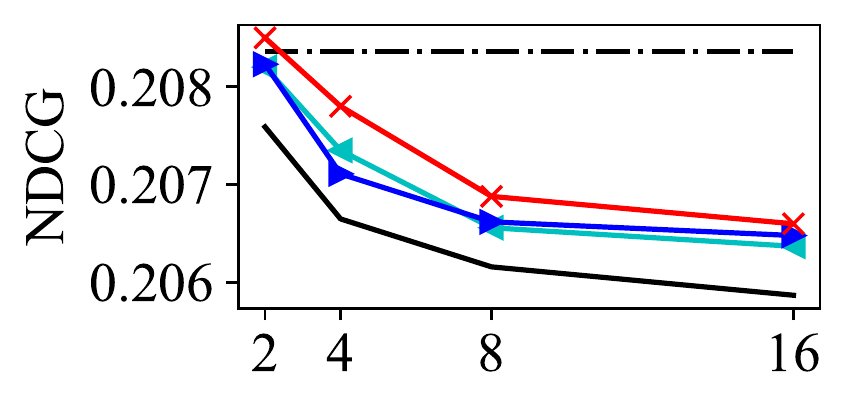}
    }
    \subfigure[DMF - rand@5]{
        \includegraphics[width=0.15\textwidth]{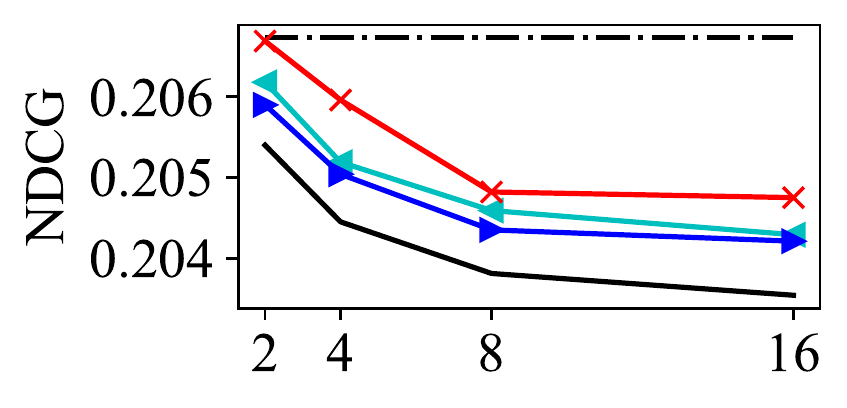}
    }
    \subfigure[NMF - rand@5]{
        \includegraphics[width=0.15\textwidth]{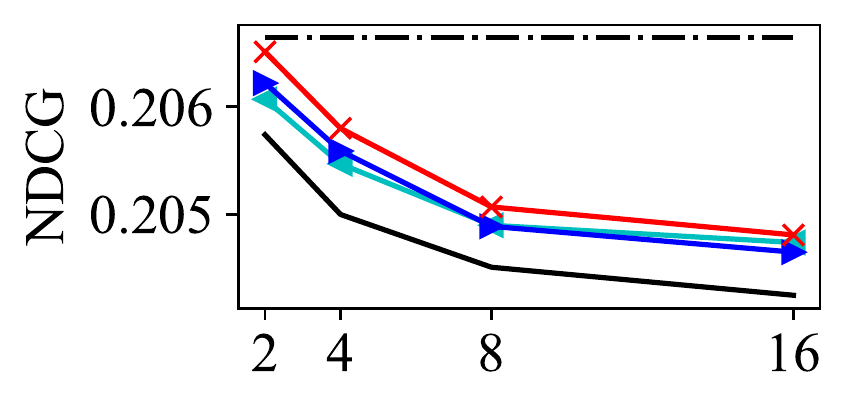}
    }
    \subfigure[DMF - top@5]{
        \includegraphics[width=0.15\textwidth]{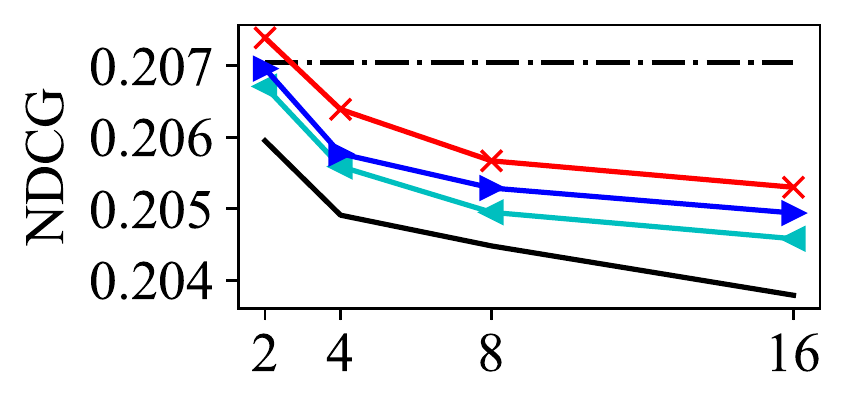}
    }
    \subfigure[NMF - top@5]{
        \includegraphics[width=0.15\textwidth]{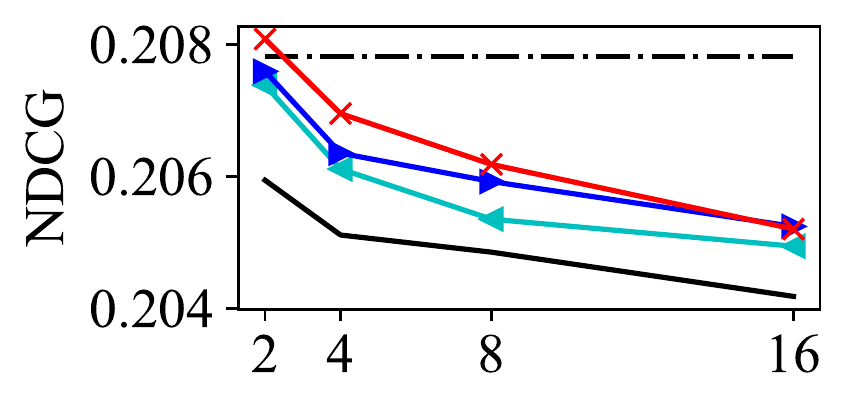}
    }
    \subfigure[DMF - learn]{
        \includegraphics[width=0.15\textwidth]{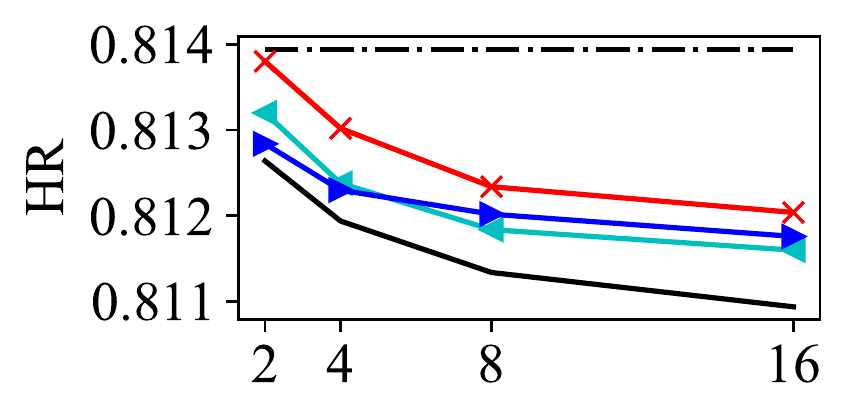}
    }
    \subfigure[NMF - learn]{
        \includegraphics[width=0.15\textwidth]{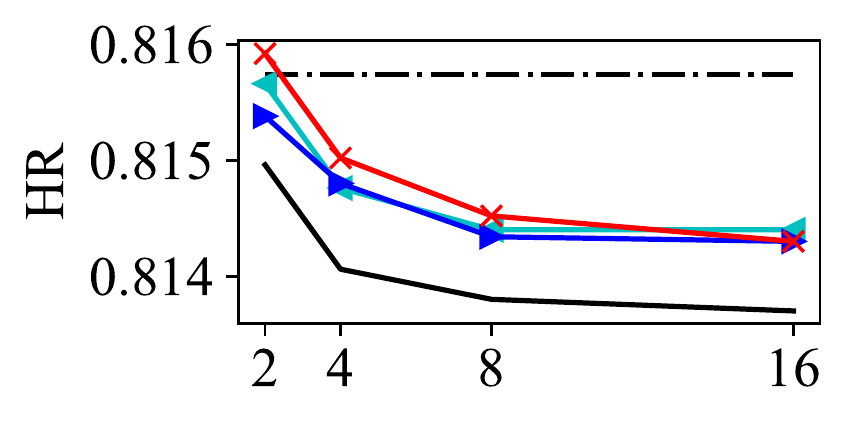}
    }
    \subfigure[DMF - rand@5]{
        \includegraphics[width=0.15\textwidth]{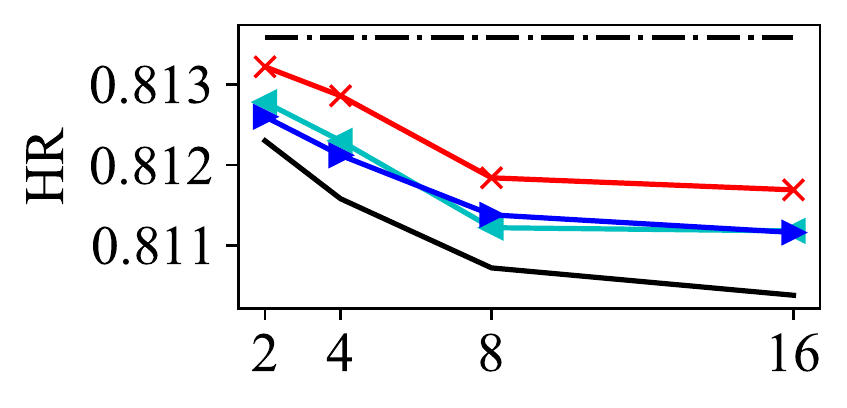}
    }
    \subfigure[NMF - rand@5]{
        \includegraphics[width=0.15\textwidth]{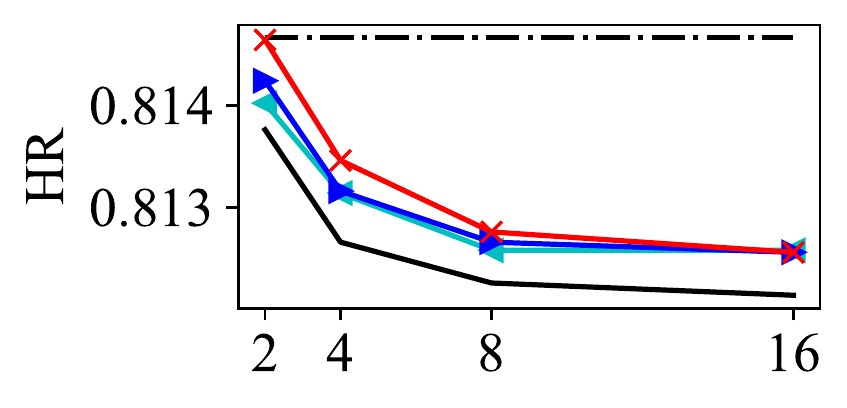}
    }
    \subfigure[DMF - top@5]{
        \includegraphics[width=0.15\textwidth]{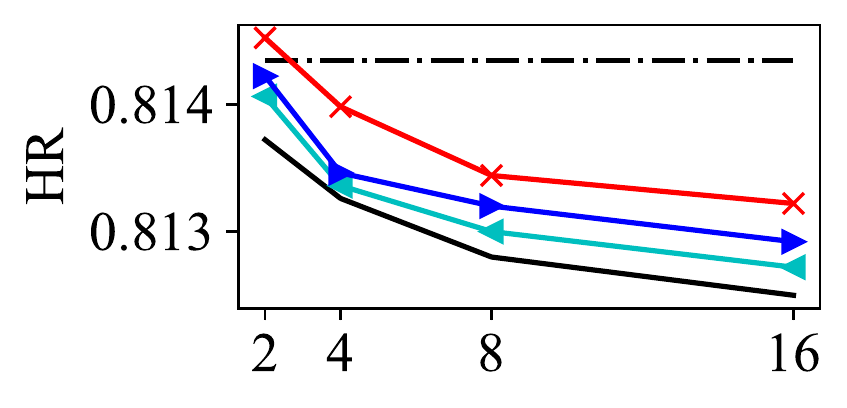}
    }
    \subfigure[NMF - top@5]{
        \includegraphics[width=0.15\textwidth]{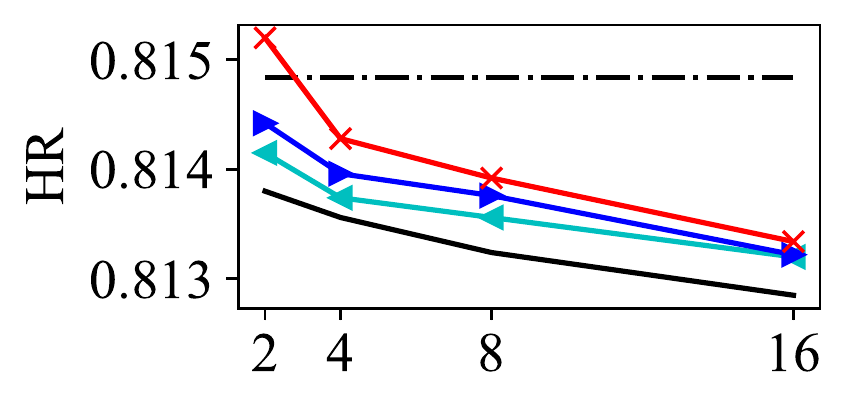}
    }
    \vspace{-0.4cm}
    \caption{NDCG@10 and HR@10 results for utility (G3) study and ablation study (Group module) on AM dataset with std under 1e-3.}
    \label{fig:metric_ad}
\end{figure*}

\subsection{Experimental Settings}

We test our proposed framework on two well-known CF models, i.e., DMF~\cite{xue2017deep} and NMF~\cite{he2017neural}. 
Specifically, we set learning rate to 0.001, total training epochs $T$ = 50 for ML dataset and 20 for AM dataset.
The dimension of user (item) embedding matrix is 16 and network structures in both DMF and NMF are set as two layers (64, 32).
We initialize the model parameters with a Gaussian distribution $\mathcal{N}(0, 0.01^2)$.
All unlearning methods for comparison are listed as follows:

\nosection{Retrain:} Retraining from scratch, i.e., $\mathcal{M}_u^*$.

\nosection{C-SISA:} We modify the State-Of-The-Art (SOTA) unlearning framework, i.e., SISA~\cite{bourtoule2021machine}, to fit recommendation setting. 
Since the original SISA trains one model on each user group in isolation, it can only learn the user embeddings within the group. 
Thus, one has to concatenate fragmentary user embedding matrices to obtain the final user embedding matrix $\alpha$.
We name this method as Concatenated-SISA and C-SISA for short.
%

\nosection{L-Rand:} \model{} via balanced random grouping.

\nosection{L-BKM:} \model{} via BKM with rating data.

\nosection{L-CBKM:} \model{} via BKM with collaborative embedding.

All models and algorithms are implemented with Python 3.8 and PyTorch 1.9.
We run all experiments on the same Ubuntu 20.04 LTS System server with 48-core CPU (Intel Xeon Gold 5118, 2.3GHz), 256GB RAM and NVIDIA GeForce RTX 3090 GPU.
We ran all models for 10 times and report the means and standard deviations (std).


\subsection{Computational Efficiency (G2)}

We evaluate the efficiency by computing the running time of three methods, i.e., Retrain, C-SISA, and L-CBKM (on behalf of \model{}).
Note that we only report the time of unlearning, because the main problem of an erasable RS is unlearning and the three methods are theoretically similar in learning time.
We randomly unlearn 5\% of users in the last group for each dataset.
In order to fully exploit the efficiency of C-SISA, we run each model in parallel for C-SISA. 
As DMF and NMF cost comparable training time, we report the running time of DMF in Figure~\ref{fig:time} for conciseness. 
We can observe consistent results in both datasets, i.e., (i) our proposed L-CBKM achieves almost comparable running time with paralleled C-SISA, and (ii) C-SISA and L-CBKM cost less unlearning time as the number of groups increases. 
The results show the efficiency of \model, especially when the group size is large which usually comes with utility loss, as we will report later.

\begin{figure*}[t]
    \centering
    \subfigure[ML-NDCG@10]{
        \includegraphics[width=0.22\textwidth]{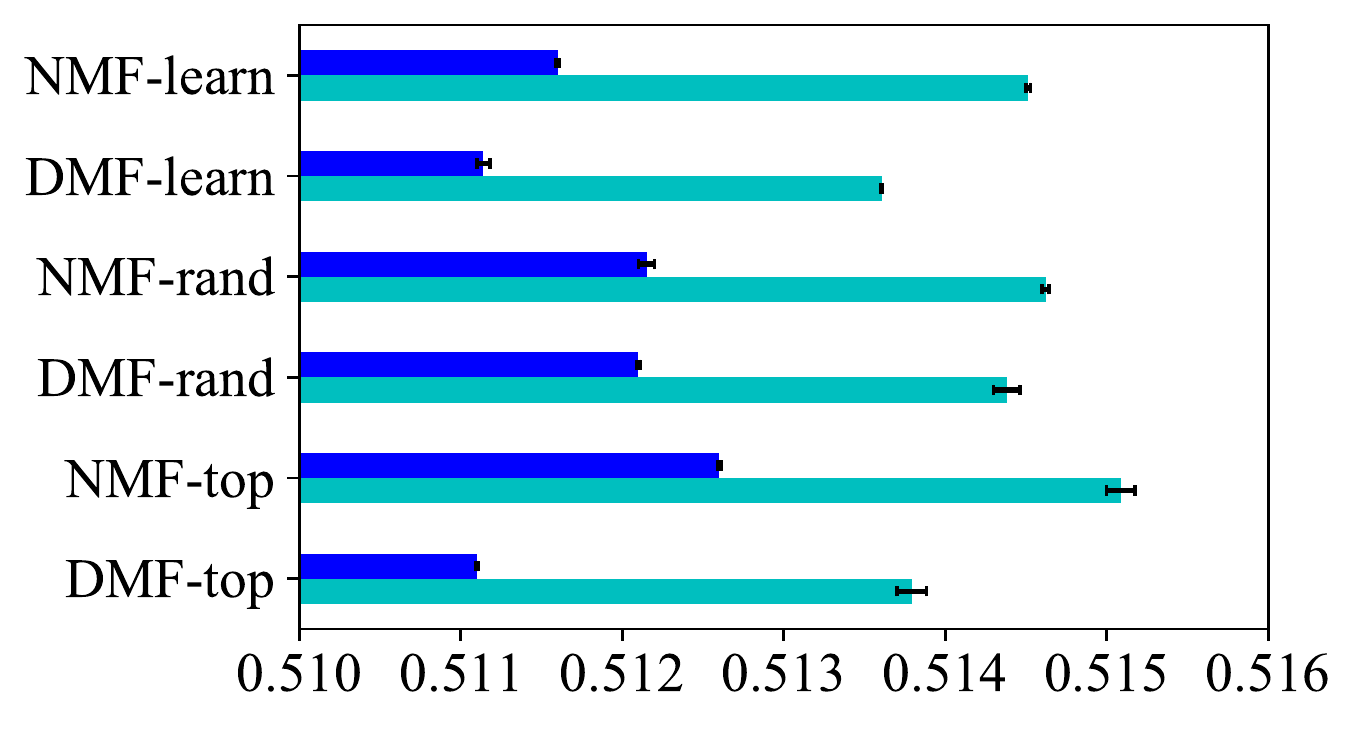}
    }
    \subfigure[ML-HR@10]{
        \includegraphics[width=0.22\textwidth]{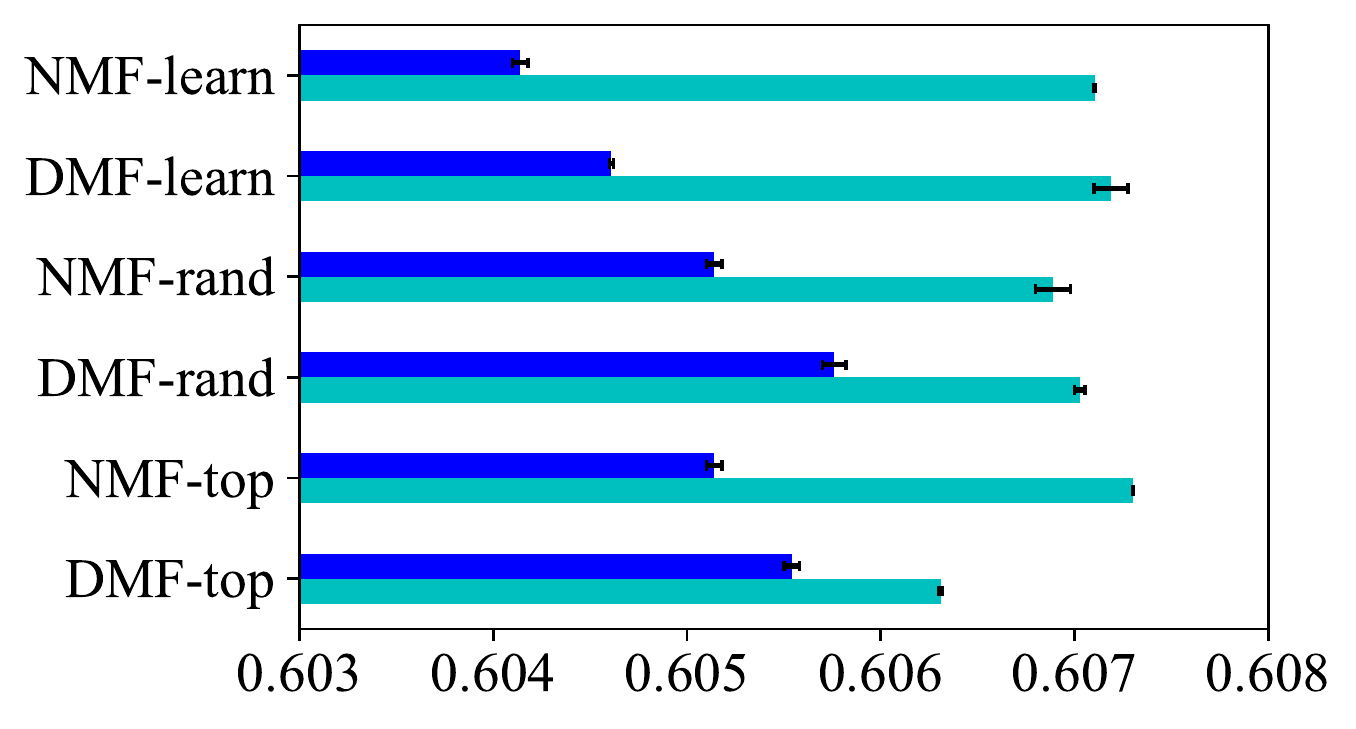}
    }
    \subfigure[AM-NDCG@10]{
        \includegraphics[width=0.22\textwidth]{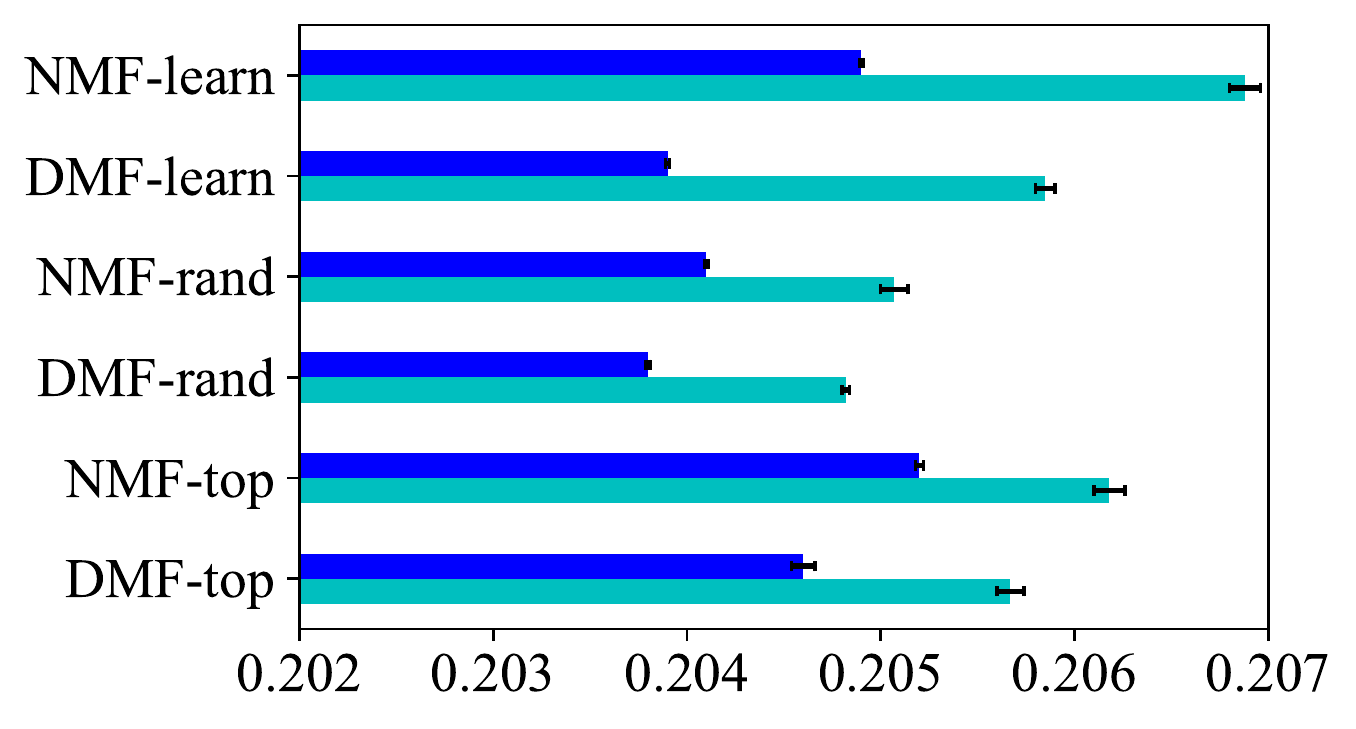}
    }
    \subfigure[AM-HR@10]{
        \includegraphics[width=0.22\textwidth]{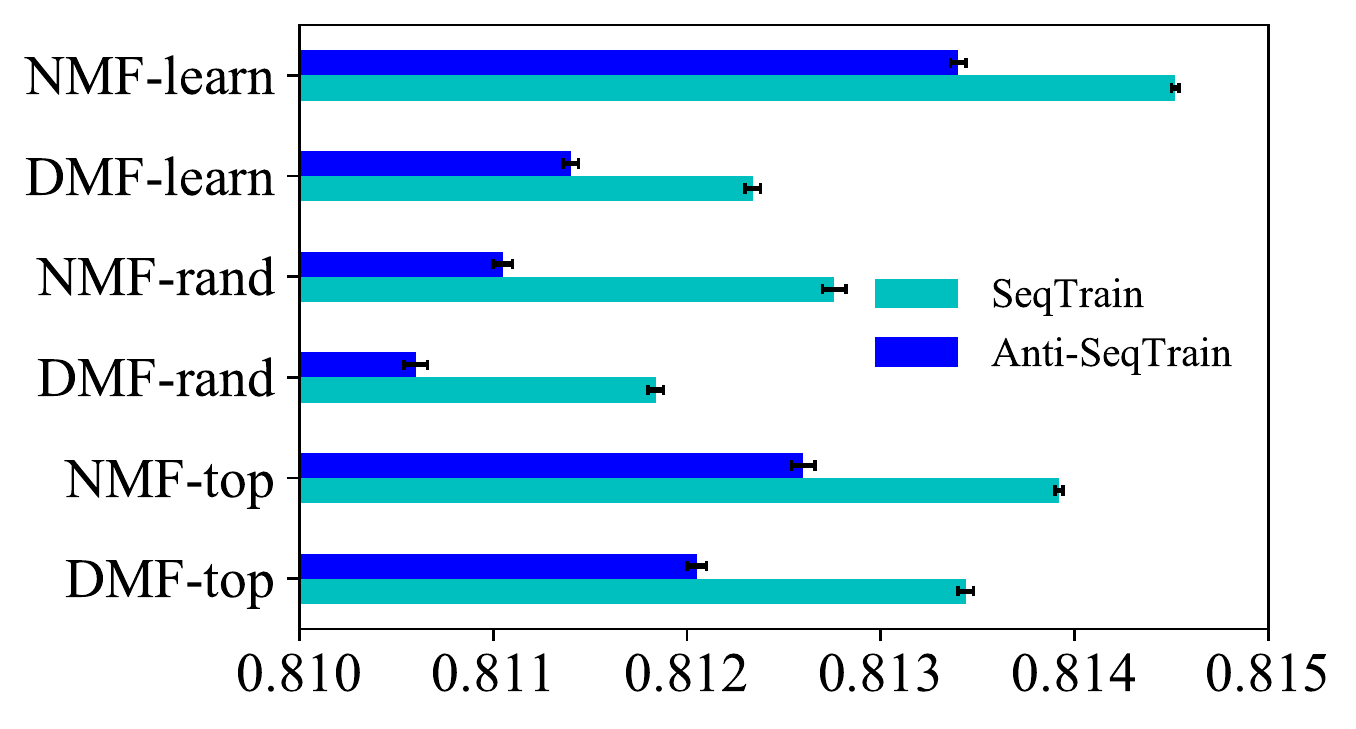}
    }
    \vspace{-0.2cm}
    \caption{Results of ablation study (SeqTrain module), where (a), (b), (c), and (d) share the same legend. }
    \label{fig:order}
    \vspace{-0.1cm}
\end{figure*}

\begin{figure}[t]
    \centering
    \subfigure[ML]{
        \label{fig:effect_ml1}
        \includegraphics[width=0.2\textwidth]{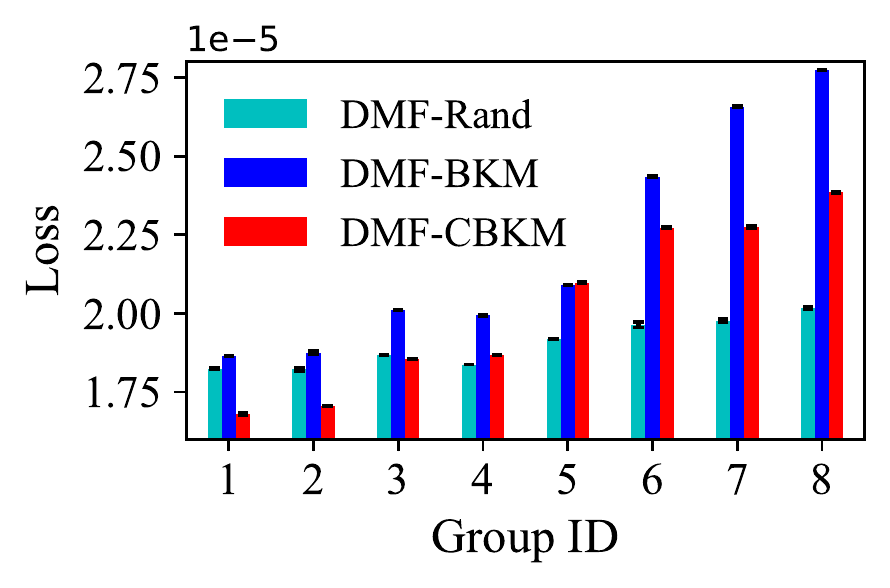}
    }
    \subfigure[AM]{
        \label{fig:effect_ad}
        \includegraphics[width=0.22\textwidth]{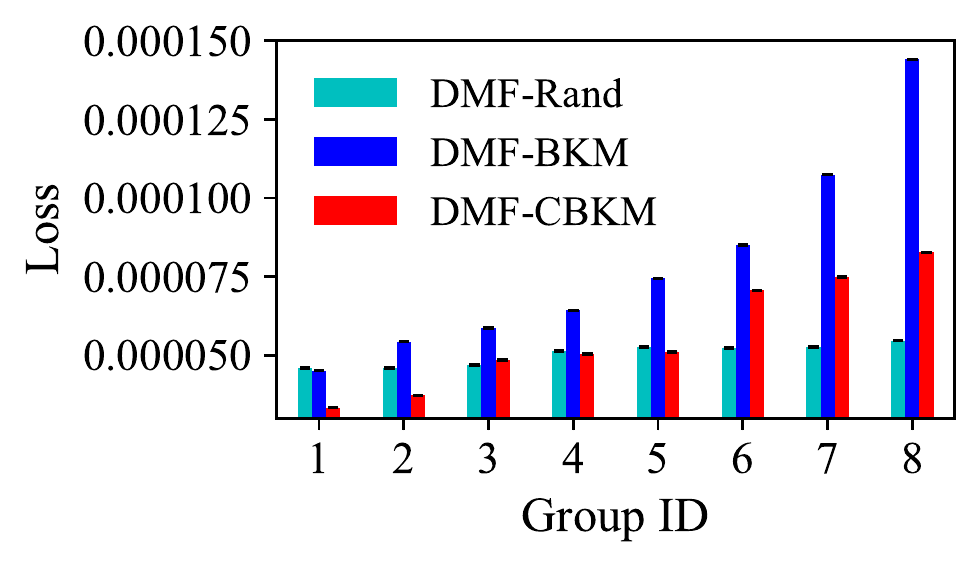}
    }
    \vspace{-0.2cm}
    \caption{Validation of collaborative \density{}. 
    }
    \label{fig:effect}
    \vspace{-0.1cm}
\end{figure}

\subsection{Model Utility (G3)}

We use two common metrics, i.e., Normalized Discounted Cumulative Gain (NDCG) and Hit Ratio (HR), to evaluate the performance of CF models~\cite{he2015trirank,xue2017deep}.
We truncate the ranked list at 10 for both metrics and report NDCG@10 and HR@10 during both learning and unlearning processes.
In order to fully study the effect of unlearning, we define two types of user-wise unlearning request, i.e., rand@$K$ and top@$K$, which denote unlearning $K\%$ random users and the top $K\%$ users (w.r.t. the number of ratings), respectively. 
We vary $K$ in $\{2.5, 5\}$ for both datasets.
%

We compare \model{} (L-Rand, L-BKM and L-CBKM) with the SOTA MU framework C-SISA, and report the results in Figure~\ref{fig:metric_ml} and \ref{fig:metric_ad}, where we vary grouping number $S$ in $\{1, 2, 4, 8, 16\}$ and set $K$ to $5$.
From them, we observe that 
(i) even if the users are randomly grouped (L-Rand), \model{} outperforms C-SISA in all testing cases, which means that, in the context of recommendation, \model{} can enhance performance by preserving collaborative information, 
(ii) the difference between Retrain and other testing methods tends to be stable as $S$ increases, which indicates that \model{} is robust with large grouping numbers,
and (iii) good performance in the learning process leads to good performance in the unlearning process.
%
Due to page limit, we report the results with $K = 2.5$ in Appendix D.1. 

\subsection{Ablation Study}\label{sec:ab}


\subsubsection{Group Module} 
We compare the effectiveness of different grouping methods (Rand, BKM, and CBKM) and report the results in Figure~\ref{fig:metric_ml} and \ref{fig:metric_ad}.
For DMF, CBKM outperforms other grouping methods in most of testing cases, while BKM achieves similar performance with Rand, because it cannot properly group the users with the sparse data.
For NMF, the results are consist with DMF, except that all grouping methods achieve similar performance when $S=16$.
This is probably because NMF's strong learning ability helps it to better capture user feature when $S$ is large, which weakens the effect of different grouping methods~\cite{rendle2020neural,xu2021rethinking}. 
%
Additionally, to further compare the effect of sparse data and collaborative embedding, we report their collaborative \density{} $\rho$ in Appendix D.2.
In summary, CBKM groups users effectively in most cases on both datasets. 

\subsubsection{SeqTrain Module}
In order to study the influence of the SeqTrain module, we verify (i) the effectiveness of training order, and (ii) the validity of collaborative \density{}.

\paragraph{Training Order}
We compare our proposed SeqTrain (easy-to-hard) with anti-SeqTrain (hard-to-easy) which trains all groups in a reverse order.
We set $S$ to 8 and $K$ to $5$ for both datasets.
Figure~\ref{fig:order} clearly shows that SeqTrain outperforms anti-SeqTrain in all testing cases, which verifies the effectiveness of our proposed SeqTrain module.
%





\paragraph{Collaborative \Density{}}
We verify the validity of $\rho$ by studying the relation between testing loss and $\rho$.
We set $S$ to 8, sort the groups in descending order according to $\rho$, and train DMF on each group from scratch for $T$ epochs.
We report the testing loss in Figure~\ref{fig:effect}. 
From it, we see that the loss generally grows with the order of groups, which means that $\rho$ is negatively related with testing loss. 
That is, the larger the collaborative \density{}, the easier the model can be trained.
In other words, $\rho$ is a valid difficulty measurer which is positively related with model utility.
The result of NMF is consistent with DMF, which is reported in Appendix D.3.
\section{Related Work}\label{sec:literature}


\paragraph{Retrain Unlearning}
This approach retrains the model on $D\backslash E$ in efficient ways, rather than from scratch.
Cao and Yang~\shortcite{cao2015towards} firstly formed the task of MU and proposed a statistical query learning based unlearning framework. 
The basic idea is to transform training samples into a reduced number of summations.
Thus, the models can retrain more efficiently on these summations.
Ginart~et~al.~\shortcite{ginart2019making} proposed efficient unlearning algorithms for $k$-means clustering with bounds of unlearning time complexity.
Lately, Bourtoule~et~al.~\shortcite{bourtoule2021machine} proposed a general unlearning framework, i.e., SISA, for most of existing machine learning models. 
SISA divides the dataset into several disjoint subsets and trains one model on each subset. 
Finally, SISA aggregates all models via majority vote.
When SISA receives an unlearning request, it only needs to retrain a subset of data. 

\paragraph{Reverse Unlearning}
This approach aims to withdraw the target data lineage from a learned model through reverse operations on $E$.
Baumhauer, Schöttle, and Zeppelzauer~\shortcite{baumhauer2020machine} applied linear filtration on corresponding parameters to unlearn the entire class for logit-based classification. 
%
%
Sekhari~et~al.~\shortcite{sekhari2021remember} focused on convex loss and proposed an unlearning algorithm based on reverse gradient operations with theoretical guarantee.
%





\section{Conclusion}

In this paper, we propose an erasable recommendation (\model{}) framework which consists of \textit{Group module} and \textit{SeqTrain module}.
The Group module partitions users into balanced groups according to their collaborative embedding.
The SeqTrain module sequentially trains all groups in an easy-to-hard learning order with \textit{collaborative \density{}} as the difficulty measurer.
%
%
Our theoretical analysis reveals that SeqTrain module can improve model utility.
Extensive experiments on two real-world recommendation datasets demonstrate that \model{} can not only achieve efficient unlearning, but also outperform the state-of-the-art unlearning models in terms of model utility.

\appendix

\section{Collaboration Embedding via Hypergraph}

There are three steps to learn collaborative embedding via hypergraph, i.e., hypergraph building, random walk to sample user sequence, and user sequence embedding.
In this section, we first introduce the preliminaries of hypergraph, then present the details of the above three steps.


\subsection{Preliminary of Hypergraph}

Agarwal~et~al.~\shortcite{agarwal2006higher} showed that the hypergraph with edge-independent vertex weight can be reduced to either clique graph or star graph.
In order to sufficiently represent the high-order relations in recommendation, we learn collaborative embedding via the hypergraph with edge-dependent vertex weight~\cite{chitra2019random}.
For conciseness, we refer to `hypergraph with edge-dependent vertex weight' as `hypergraph' in the rest of this paper.

A hypergraph can be formulated as $\mathcal{G} = \{\mathcal{V}, \mathcal{E}\, \mathcal{W}\}$, where $\mathcal{V}$ denotes the vertex set $\{v_i\}$, $\mathcal{E}$ denotes the hyperedge set $\{e_i\}$ ($e_i \in 2^\mathcal{V}$), and $\mathcal{W}$ denotes the weight set $\{w(i, j) | \text{for\enspace} \forall i, j \text{\enspace that\enspace} e_i \in \mathcal{E}, v_j \in e_i\}$.

\subsection{Hypergraph Building}

In order to build a hypergraph, we need association rules to connect a hypergraph with the original bipartite graph which is equivalent to the user-item interaction matrix.
There are three rules for building vertex, hyperedge, and weight respectively.

\paragraph{Vertex Building Rule} We define each user as a vertex in hypergraph, which means $|\mathcal{V}| = N$.

\begin{definition}[User's $l$-order reachable neighbors]
    In a user-item bipartite graph, $user_i (user_j)$ is $user_j (user_i)$'s l-order reachable neighbor if there exists a sequence of adjacent vertices, i.e., path, between $user_i$ and $user_j$, and the number of vertices in this path is smaller than $l$. 
\end{definition}

\paragraph{Hyperedge Building Rule} 
We define the user's $l$-order reachable neighbors, which is represented as a hyperedge in hypergraph, to model the high-order relations.
Correspondingly, the users, i.e., neighbors, within the neighborhood are vertices that connected by this hyperedge.
Following \cite{ji2020dual}, we set $l=4$ in this paper.

\paragraph{Weight Building Rule} We define $w(i, j)$ as the average rating of $user_j$ within $user_i$'s $l$-order reachable neighbors.

\subsection{Random Walk}

\begin{algorithm}[tb]
\caption{Random Walk on Hypergraph}
\label{alg:rw}
\textbf{Input}: Hypergraph: $\mathcal{G}$, walk repetition: $rep$, walk depth: $dep$\\
\textbf{Output}: User sequences: $F \in \mathbb{R}^{(N\cdot rep)\times dep}$\\
\textbf{Procedure}:
\begin{algorithmic}[1] 
\FOR{$v_i$ in $\mathcal{V}$}
    \FOR{$j$ in $[rep]$}
        \STATE Start a user sequence with $v_i$, $v_{i, 1} = v_i$;
        \FOR{$k$ in $[dep]$}
            \STATE Select a hyperedge $e$ containing $v_{i, k}$ with probability proportional to $|e|$;
            \STATE Select a vertex $v$ from $e$ with probability proportional to $w(e, v)$;
            \STATE Add $v$ into user sequence, $v_{i, k+1} = v$;
        \ENDFOR
        \STATE Complete the user sequence and add it to $F$;
    \ENDFOR
\ENDFOR
\end{algorithmic}
\end{algorithm}

A random walk on a hypergraph is typically defined as follows~\cite{chitra2019random}.
We modify the traditional random walk to suit recommendation setting and present it in Algorithm~\ref{alg:rw}.
Our proposed random walker migrates back and forth between vertices and hyperedge to sample user sequences.
From a vertex, the walker choose a hyperedge based on its size, i.e., number of vertices it connects (line 5 in Algorithm~\ref{alg:rw}).
From a hyperedge, the walker choose a vertex based on its associated weight (line 6 in Algorithm~\ref{alg:rw}).

Obviously, walk repetition and walk depth are two hyper-parameters in random walk.
According to our empirical study, we set walk repetition as 4 and walk depth as 8 in this paper.

\subsection{Sequence Embedding}

After random walk, we have $N\cdot rep$ user sequences. 
Regarding the sequence of users as the sentence of words, we can apply sequence embedding techniques in natural language processing field to learn the collaborative embedding.
In this paper, we apply the widely used Word2Vec~\cite{church2017word2vec} to learn the embedding from sampled user sequences.
The choice of sequence embedding techniques is not a focus of this paper, please refer to the original paper for more details.

\section{Balanced Grouping Method}

\subsection{Unbalanced Phenomenon}

\begin{figure}[t]
    \includegraphics[width=\linewidth]{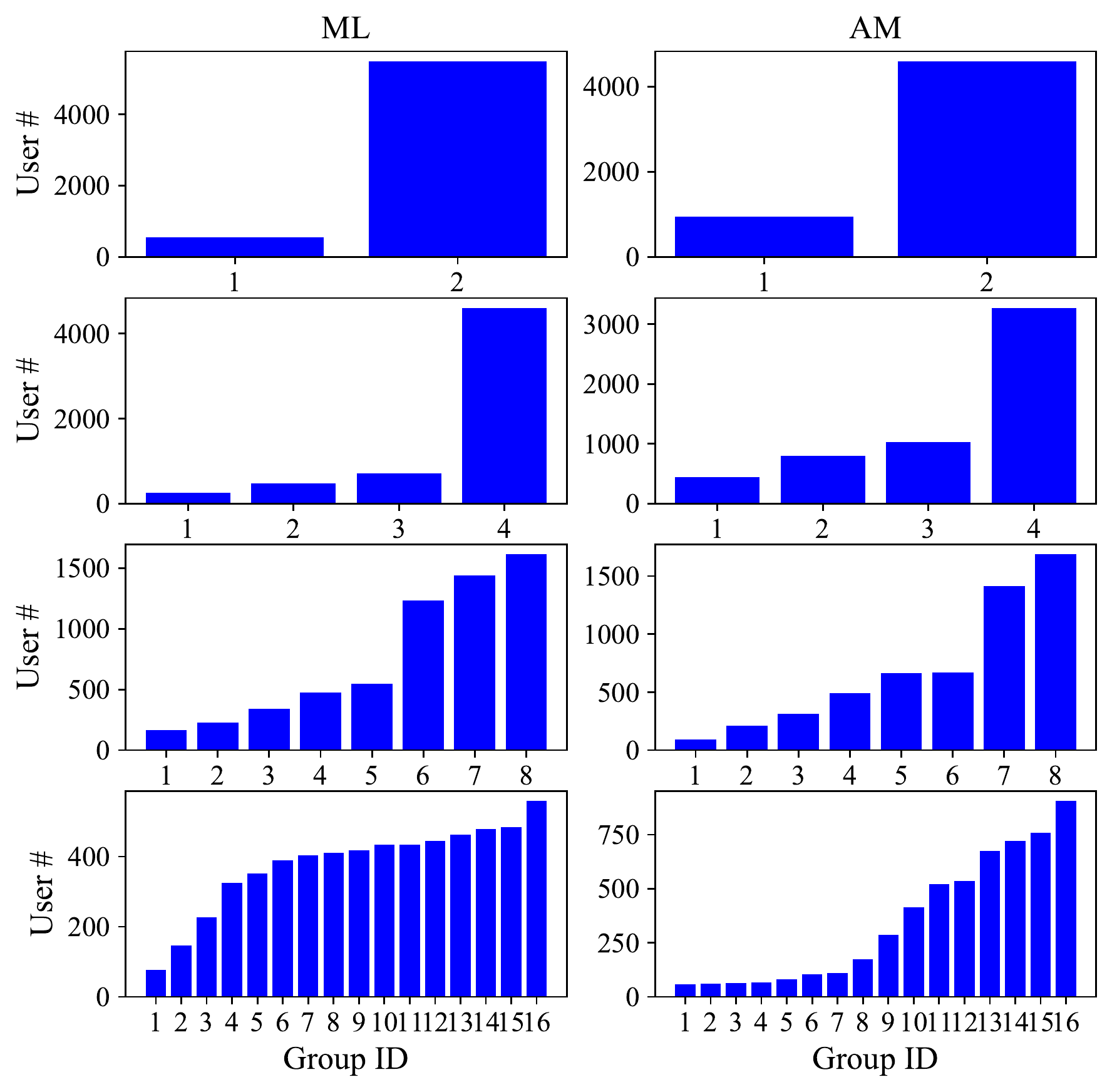}
    \caption{Distributions of group size (user number within the group) via $k$-means clustering on two datasets, i.e., MovieLens 1M (ML) and Amazon Digital Music (AM), with $k = 2, 4, 8, 16$.}
    \label{fig:unbalance}
\end{figure}

We conduct $k$-means clustering on two real-world datasets with $k = 2, 4, 8, 16$ and report the result in Figure~\ref{fig:unbalance}.
We find that the grouping distributions are highly unbalanced on both datasets.

\subsection{Efficiency Analysis}

We use $c_i$ to denote the training time cost of $i$-th group which is in proportion to the group size.
Our proposed LASER adopts a sequential training manner, therefore the retraining time cost from $i$-th group is $\sum_{j=i}^S c_j$.
Since we assume that users submit unlearning requests with equal probability, the probability of a request locating in $i$-th group is $c_i/Z$ where $Z = \sum_i c_i$.
Thus, the expectation of retraining time cost can be written as:
\begin{equation}\label{equ:ori}
    \mathbb{E}(C) = \sum_{i=1}^S\Big((\sum_{j=i}^S c_j) c_i / Z\Big).
\end{equation}

We prove that $\mathbb{E}(C)$ reaches the minimum when $c_i = Z/n$, which means balanced grouping can maximize the unlearning efficiency.

\begin{proof}
    Based on Multinomial Theorem
    , we can rewrite (\ref{equ:ori}) as:
    \begin{align}\label{equ:mod}
        \mathbb{E}(C) & = \frac{1}{2Z}\Big((\sum_{i=1}^n c_i)^2 + \sum_{i=1}^n c_i^2\Big)\nonumber\\
        & = \frac{1}{2Z}\Big(Z^2 + \sum_{i=1}^n c_i^2\Big).
    \end{align}
    According to Cauchy-Schwarz Inequality~\cite{bhatia1995cauchy}, for $n$ random variables $x_i, y_i$, we have:
    \begin{equation}
        \sum_{i=1}^n x_i^2 \sum_{i=1}^n y_i^2 \ge (\sum_{i=1}^n x_i y_i)^2.
    \end{equation}
    Setting $y_i = 1$ for all $i$, we can get a lower bound of $\sum_i x_i^2$ as:
    \begin{equation}
        \sum_{i=1}^n x_i^2 \ge \frac{(\sum_{i=1}^n x_i)^2}{n}.
    \end{equation}
    Taking it into (\ref{equ:mod}), we have:
    \begin{equation}
        \mathbb{E}(C) \ge \frac{Z}{2}(1 + \frac{1}{n})
    \end{equation}
    We can easily compute that setting $c_i = Z/n$ achieves this lower bound of $\mathbb{E}(C)$.
\end{proof}

\subsection{Balanced $k$-means Clustering}

Our proposed balanced grouping method (Algorithm~2) can be applied to a wide range of clustering algorithms. 
We take balanced $k$-means clustering as an example and present the details in Algorithm~\ref{alg:sim}.

\begin{algorithm}[tb]
\caption{$ComputeSimilarity$ ($k$-means version)}
\label{alg:sim}
\textbf{Input}: User embedding matrix: $B$, group label: $\Lambda$\\
\textbf{Output}: Priority list $P$\\
\textbf{Procedure}:
\begin{algorithmic}[1] 
\STATE $P =: $ empty list;
\STATE compute $N$ centroids with respect to $\Lambda$;
\FOR{$(i, j)$ in all user-centroid pairs}
    \STATE append $dist(i, j)$ to $P$;
\ENDFOR
\end{algorithmic}
\end{algorithm}

\section{Proof of Theorem~\ref{the:p}}

\begin{theorem}\label{the:p}
    Given a Bayesian prior $p$ for the model parameters $\theta$, we have:
    \begin{equation}
        \mathcal{U}_p(\theta) = \mathcal{U}(\theta) + \hat{\text{Cov}}[U_\theta, p].
    \end{equation}
\end{theorem}

\paragraph{Recall} Let $\theta$ denotes the model parameter vector and $L_\theta(g_i)$ denotes the loss of the model of $i$-th group. 
Adopting empirical risk minimization framework, we have the empirical loss
\begin{align}
    \mathcal{L}(\theta) = \hat{\mathbb{E}}[L_\theta] = \frac{1}{S}\sum_{i = 1}^S L_\theta(g_i), \enspace\theta^* = \underset{\theta}{\operatorname{\arg\min}}\mathcal{L}(\theta).
\end{align}
Minimizing the empirical loss can be regarded as maximizing model utility~\cite{hacohen2019power}, which is defined as:
\begin{align}\label{equ:u}
    \mathcal{U}(\theta) & = \hat{\mathbb{E}}(U_\theta) = \frac{1}{S}\sum_{i = 1}^S U_\theta(g_i) \triangleq \frac{1}{S}\sum_{i = 1}^S e^{-L_\theta(g_i)}, \nonumber\\
    \theta^* & = \underset{\theta}{\operatorname{\arg\max}}\hspace{1mm}\mathcal{U}(\theta).
\end{align}

\begin{proof}
    With the help of the Bayesian prior $p$, we can formulate (\ref{equ:u}) as:
    \begin{align}\label{equ:up}
        \mathcal{U}_p(\theta) & = \hat{\mathbb{E}}_p[U_\theta] = \sum_{i=1}^S U_\theta(g_i)p_i = \sum_{i=1}^S e^{-L_\theta(g_i)}p_i, \nonumber\\
        \enspace\theta^* & = \underset{\theta}{\operatorname{\arg\max}}\hspace{1mm}\mathcal{U}_p(\theta),
    \end{align}
    where $p_i$ denotes the prior probability of $i$-th group.
    As $U_\theta$ and $p$ are two random variables in $\mathcal{U}_p(\theta)$, we can rewrite (\ref{equ:up}) as follows:
    \begin{align}
        \mathcal{U}_p(\theta) & = S\hat{\mathbb{E}}[U_\theta]\hat{\mathbb{E}}[p] + \sum_{i=1}^S (U_\theta(g_i) - \hat{\mathbb{E}}[U_\theta])(p_i - \hat{\mathbb{E}}[p])\nonumber\\
        & = \mathcal{U}(\theta) + \hat{\text{Cov}}[U_\theta, p].
    \end{align}
\end{proof}

\section{More Empirical Results}

\subsection{Utility (G3)}

\begin{figure}[t]
    \centering
    \includegraphics[width=0.45\textwidth]{fig/metirc_legend.pdf}\\
     \vspace{-2mm}
    \subfigure[DMF - rand@2.5]{
        \includegraphics[width=0.2\textwidth]{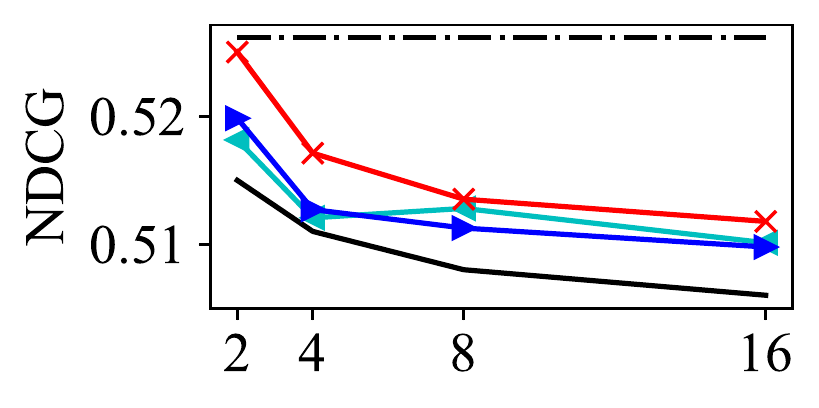}
    }
    \subfigure[NMF - rand@2.5]{
        \includegraphics[width=0.2\textwidth]{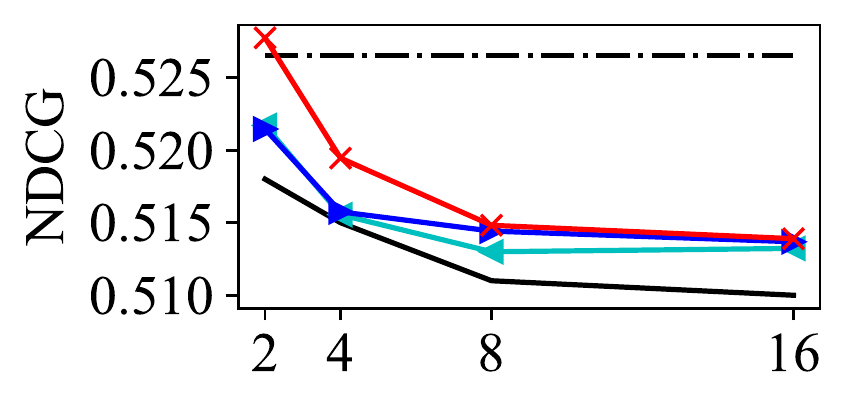}
    }
    \subfigure[DMF - top@2.5]{
        \includegraphics[width=0.2\textwidth]{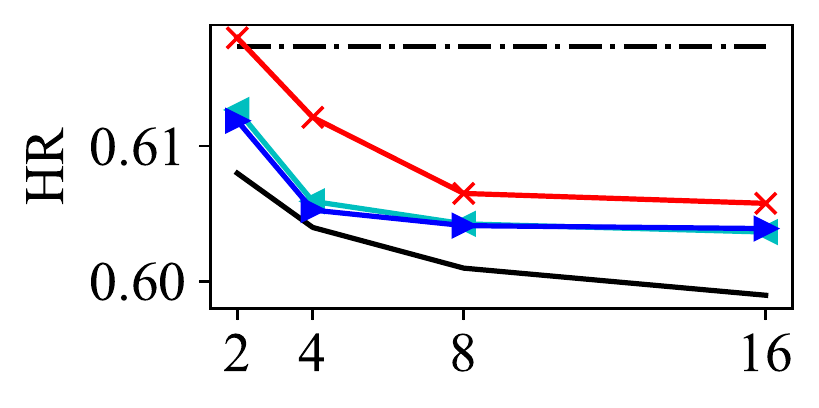}
    }
    \subfigure[NMF - top@2.5]{
        \includegraphics[width=0.2\textwidth]{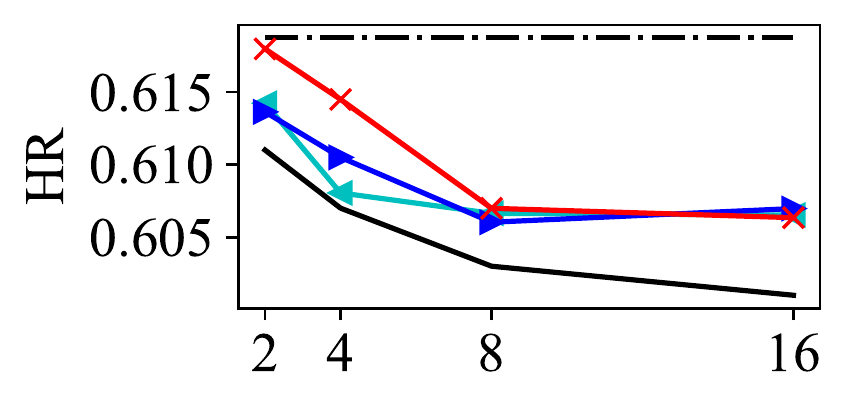}
    }
    \vspace{-4mm}
    \caption{NDCG@10 and HR@10 results for ML dataset with std under 1e-3.}
    \label{fig:metric_ml_apd}
    \vspace{-2mm}
\end{figure}

\begin{figure}[t]
    \centering
    \includegraphics[width=0.45\textwidth]{fig/metirc_legend.pdf}\\
    \vspace{-2mm}
    \subfigure[DMF - rand@2.5]{
        \includegraphics[width=0.2\textwidth]{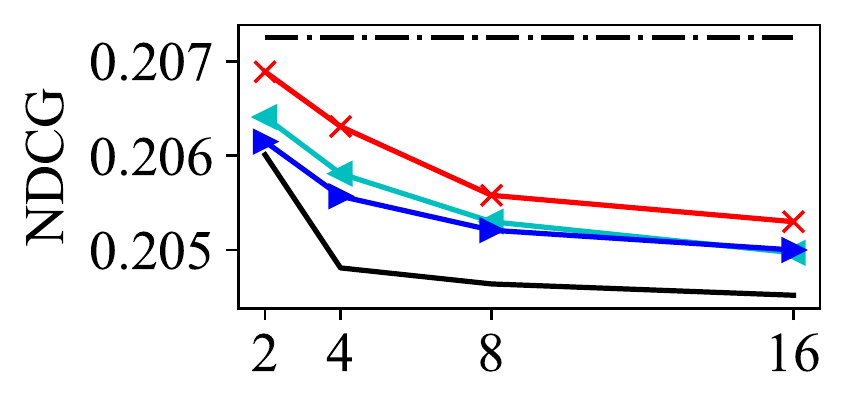}
    }
    \subfigure[NMF - rand@2.5]{
        \includegraphics[width=0.2\textwidth]{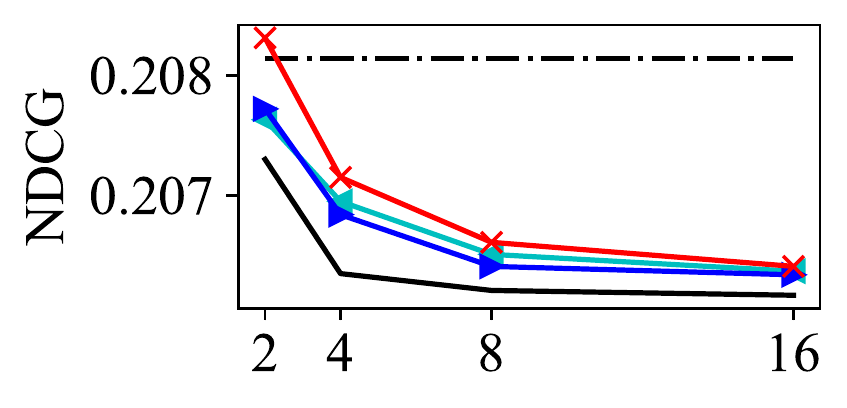}
    }
    \subfigure[DMF - top@2.5]{
        \includegraphics[width=0.2\textwidth]{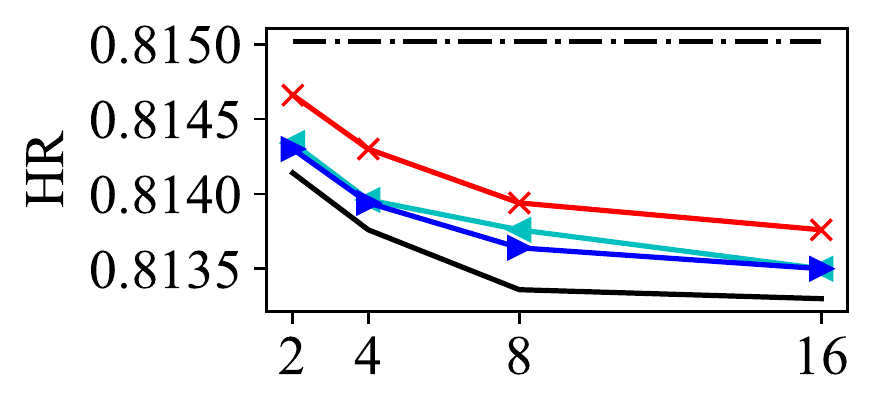}
    }
    \subfigure[NMF - top@2.5]{
        \includegraphics[width=0.2\textwidth]{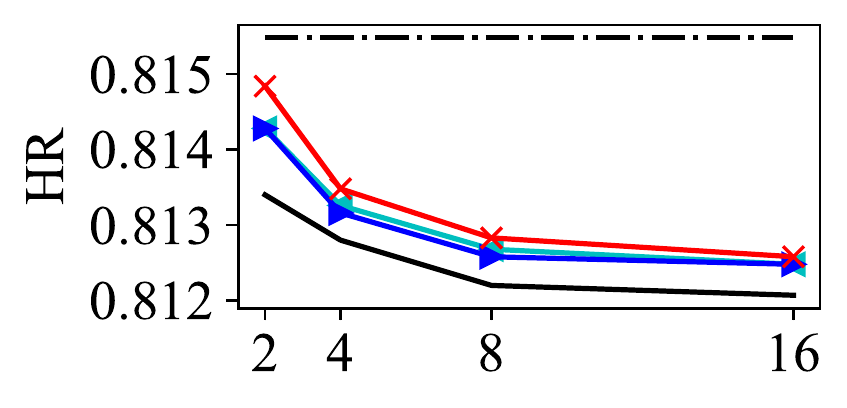}
    }
    \vspace{-4mm}
    \caption{NDCG@10 and HR@10 results for AM dataset with std under 1e-3.}
    \label{fig:metric_ad_apd}
    \vspace{-2mm}
\end{figure}

We report the results of utility metrics with two unlearning request types (rand@2.5 and top@2.5) in Figure~\ref{fig:metric_ml_apd} and Figure \ref{fig:metric_ad_apd}.
We observe that the results are consistent with that of rand@5 and top@5.

\subsection{Comparison of Sparse Data and Collaborative Embedding}

\begin{figure}[t]
    \centering
    \subfigure[ML]{
        \includegraphics[width=0.22\textwidth]{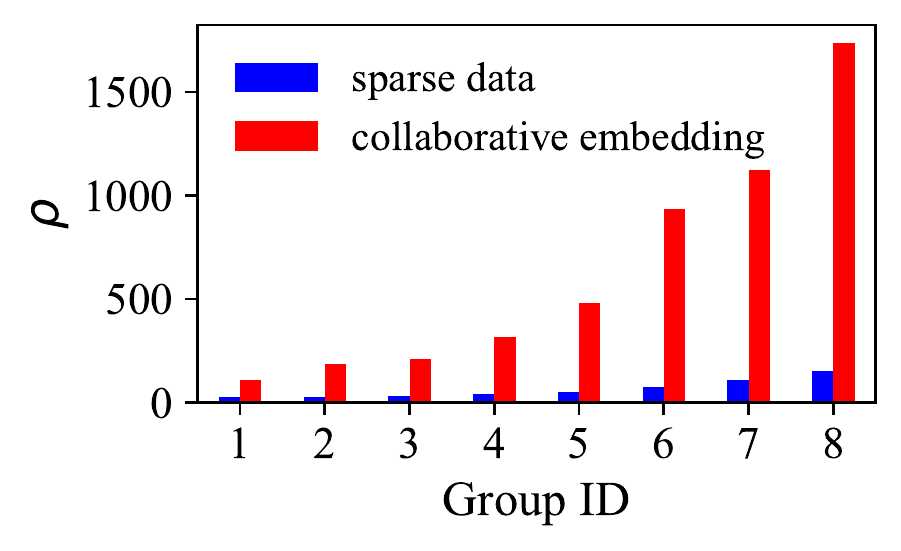}
    }
    \subfigure[AM]{
        \includegraphics[width=0.22\textwidth]{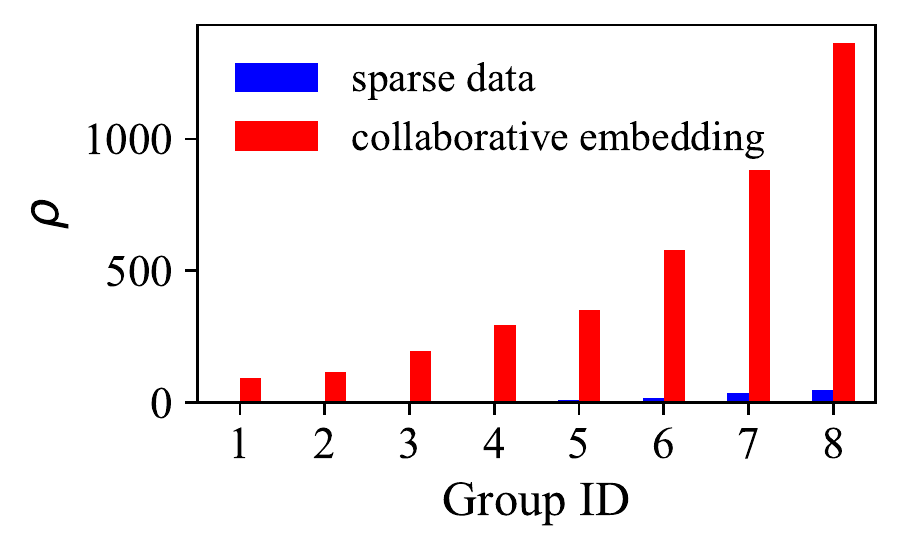}
    }
    \vspace{-0.2cm}
    \caption{Comparison of sparse data and collaborative embedding. 
    }
    \label{fig:com}
    \vspace{-0.1cm}
\end{figure}

We compare the effect of sparse data (BKM) and collaborative embedding (CBKM) in terms of collaborative \density and report the result in Figure~\ref{fig:com}.
We find that our proposed collaborative embedding can achieve much higher collaborative \density than the original sparse data.

\subsection{Verification of Collaborative \Density}
 
To verify the validity of $\rho$, we report the testing loss of NMF in Figure~\ref{fig:effect}.
We observe the consistent results with DMF.
\begin{figure}[t]
    \centering
    \subfigure[ML]{
        \label{fig:effect_ml1_apd}
        \includegraphics[width=0.2\textwidth]{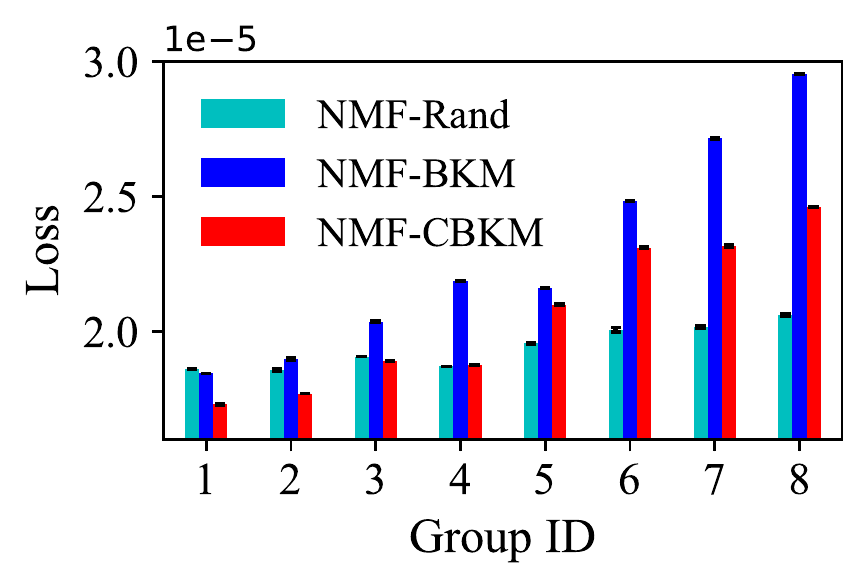}
    }
    \subfigure[AM]{
        \label{fig:effect_ad_apd}
        \includegraphics[width=0.22\textwidth]{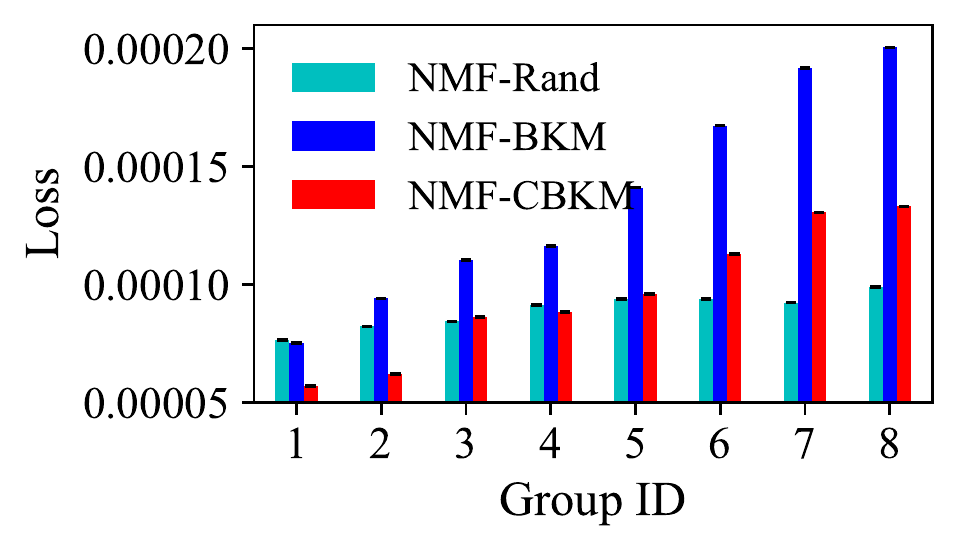}
    }
    \vspace{-4mm}
    \caption{Validation of collaborative \density{}.}
    \label{fig:effect_apd}
\end{figure}







\bibliographystyle{ijcai22}
\bibliography{ijcai22}

\end{document}